\documentclass[prx,aps,groupedaddress,twocolumn,superscriptaddress,longbibliography]{revtex4-1}
\usepackage{graphicx}
\usepackage{amsmath}
\usepackage{amssymb}

\usepackage[dvipsnames]{xcolor}
\usepackage{changepage}
\usepackage{tcolorbox}
\usepackage{pifont}
\usepackage{enumitem}
\usepackage{multirow}
\usepackage[export]{adjustbox}
\usepackage{braket}

\newlist{todolist}{itemize}{2}
\setlist[todolist]{label=$\square$}

\usepackage[colorlinks=true,linkcolor=blue]{hyperref}
\usepackage{cleveref}
\hypersetup{
	colorlinks,
	linkcolor={blue!75!black},
	citecolor={blue!75!black},
	urlcolor={blue!75!black},
}

\newcommand{\Okin}{\mathcal{O}_{\mathrm{kin}}}
\newcommand{\Opot}{\mathcal{O}_{\mathrm{pot}}}
\newcommand{\Okins}{\mathcal{O}_{\mathrm{kin},\square}}
\newcommand{\Opots}{\mathcal{O}_{\mathrm{pot},\square}}

\begin{document}
\title{Sublattice scars and beyond in two-dimensional $U(1)$ quantum link lattice gauge theories}

\author{Indrajit Sau}
\affiliation{School of Physical Sciences, Indian Association for the Cultivation of Science, Jadavpur, Kolkata 700032, India}

\author{Paolo Stornati}
\affiliation{ICFO-Institut de Ciencies Fotoniques, The Barcelona Institute of Science and Technology, 
Mediterranean Technology Park, Avinguda Carl Friedrich Gauss, 3, 08860 Castelldefels, Barcelona, Spain}

\author{Debasish Banerjee}
\affiliation{Theory Division, Saha Institute of Nuclear Physics, 1/AF Bidhannagar, Kolkata 700064, India}
\affiliation{Homi Bhabha National Institute, Training School Complex,Anushaktinagar, Mumbai 400094, India}

\author{Arnab Sen}
\affiliation{School of Physical Sciences, Indian Association for the Cultivation of Science, Jadavpur, Kolkata 700032, India}

\date{\today}

\begin{abstract}
  In this article, we elucidate the structure and properties of a class of anomalous 
  high-energy states of matter-free $U(1)$ quantum link gauge theory Hamiltonians using 
  numerical and analytical methods. Such anomalous states, known as \emph{quantum many-body 
  scars} in the literature, have generated a lot of interest due to their athermal nature. 
  Our starting Hamiltonian is $H = \Okin + \lambda \Opot$, where $\lambda$ is a real-valued 
  coupling, and $\Okin$ ($\Opot$) are summed local diagonal (off-diagonal) operators in the
  electric flux basis acting on the elementary plaquette $\square$. The spectrum of the model 
  in its spin-$\frac{1}{2}$ representation on $L_x \times L_y$ lattices reveal the existence 
  of \emph{sublattice scars}, $\ket{\psi_s}$, which satisfy $\Opots \ket{\psi_s} =\ket{\psi_s}$ 
  for all elementary plaquettes on one sublattice and $\Opots \ket{\psi_s} =0$ on the other, 
  while being simultaneous zero modes or nonzero integer-valued eigenstates of $\Okin$. We 
  demonstrate a ``triangle relation'' connecting the sublattice scars with nonzero integer 
  eigenvalues of $\Okin$ to particular sublattice scars with $\Okin=0$ eigenvalues. 
  A fraction of the sublattice scars have a simple description in terms of emergent short 
  singlets, on which we place analytic bounds. We further construct a long-ranged parent 
  Hamiltonian for which all sublattice scars in the null space of $\Okin$ become unique 
  ground states and elucidate some of the properties of its spectrum. In particular, zero 
  energy states of this parent Hamiltonian turn out to be exact scars of \emph{another} 
  $U(1)$ quantum link model with a staggered short-ranged diagonal term. 
\end{abstract}

 \maketitle
 \tableofcontents

\section{Introduction}
Traditional high-energy physics has focussed on the consideration of physical phenomena which 
typically happens at energy scales much greater than the ground state, or even the relevant 
low-energy physics. This approach has been particularly useful in decoding the fundamental 
particles and their interactions in Nature through a series of collider experiments at successively 
higher energies, culminating in the discovery of the Higgs particle \cite{RevPP22}. The nature 
of the collider experiments is such that it is able to create matter at high energy densities, 
while having less control on the microscopic details of particular (eigen)states of interesting 
theories. This makes it extremely difficult to study the physics of isolated excited quantum 
states which may have interesting properties by themselves. Probing fragile quantum mechanical 
effects directly from collider experiments, such as entanglement, can be challenging, although 
there is recent work on studies of entanglement of particle pairs produced at the Large Hadron 
Collider (LHC) \cite{Ramos:2020kaj,Fabbrichesi:2021npl}.

 In the past decade, the landscape of experiments available have highly expanded due to the remarkable 
success of tabletop experiments in controlling microscopic degrees of freedom in a very precise manner
\cite{Preskill2018}. This has led to the construction of \emph{quantum simulators} and \emph{quantum computers} 
using a host of different architectures such as Rydberg atoms, superconducting qubits, and even photons 
trapped in a cavity (circuit QED) \cite{Tacchino2019}. These experiments have enabled the use of measures 
of quantum entanglement, and more generally quantum information theory, in order to classify and understand 
a variety of physical phenomena which defy conventional wisdom. Therefore, dynamical aspects of various 
theories, notably those relating to thermalization, can be studied with these tabletop experiments 
considerably easily than in collider experiments. There has been proposals to mimic particle collisions
in tabletop experiments to study their dynamics \cite{Farrell:2023fgd,Belyansky:2023rgh}, as well as
various other aspects of the physics which are difficult to study using classical simulation methods
\cite{Banuls:2019bmf,Bauer:2023qgm,Heitritter:2022jik,Banerjee:2021tfo,Halimeh:2022mct,Atas:2022dqm,
DiMeglio:2023nsa}. Together with the original proposals, they have motivated the realization of 
lattice gauge theories in quantum computer and simulator setups 
\cite{Martinez:2016yna,Mil2020,Yang2020,Klco:2018kyo}.

  Questions about the thermalization of many body quantum systems is also a central topic of investigation
in condensed matter physics. The studies of thermalization in quantum spin and fermionic models have been
guided by the so-called \emph{eigenstate thermalization hypothesis} (ETH) 
\cite{Deutsch1991,Srednicki1994,Rigol2008}. ETH postulates that even though quantum mechanics allows for 
unitary dynamics of closed systems, a subsystem of the full system \emph{appears} to thermalize, since the 
rest of the system acts as a heat bath for the subsystem under consideration, thereby establishing the 
validity of quantum statistical mechanics. A key result that has emerged from studies of ETH in the most 
common spin and fermionic model is that a state whose energy density is $O(J)$, $J \sim 1$ higher than the 
ground state thermalizes in time $t \sim O(1)$. 
 
  Simultaneously, there has also been an explosion of interest towards the investigation of scenarios in
translationally invariant systems which show deviation from the conventional wisdom of ETH, both theoretically
and experimentally~\cite{Turner2018a, Turner2018b, StarkMBL2019, Salaetal2020, Khemanietal2020, Bernien2017, 
Morong2021}. Two such scenarios, \emph{weak ergodicity breaking} and \emph{strong ergodicity breaking}, have 
been theoretically proposed for translational invariant systems. The former scenario shows the presence of 
anomalous high-energy states, also dubbed as \emph{quantum many-body scars}~\cite{Turner2018a,
Turner2018b, Vafek2017,Choi2019,Ho2019,Lin2019,Iadecola2019, Shiraishi2019,Moudgalya2018a,Moudgalya2018b,Pakrouski2020,Mark2020,Shibata2020,Sinha2020,Zhao2020,Pakrouski2021,Mukherjee2020a,
Mukherjee2020b,Mukherjee2020c,schindler2021exact,mukherjee2021periodically,Mukherjee2021spin1,McClarty2020,Karle2021} in the 
spectrum of the Hamiltonian which often admit description in terms of an exponentially smaller number of 
Fock states (compared to neighboring non-anomalous high-energy states) with a simple physical interpretation. 
The latter scenario on the other hand, emerges when the symmetry resolved Hamiltonian further splits into
exponentially many disconnected sectors without a clear description in terms of symmetries and no single 
dominant sector in the thermodynamic limit~\cite{Salaetal2020, Khemanietal2020, Mukherjee2021,anwesha2023}.
The reader is referred to Ref.~\cite{Moudgalya_2022review} for a detailed review of the topics. 
  
 Lattice gauge theories, which possess a local invariance, and thus often used as microscopic models to 
explain a variety of phenomena in both high-energy and condensed matter physics, such as confinement 
\cite{Chandrasekharan1997,Banerjee2013,Wiese2014}, spin-liquids \cite{Hermele2004}, and superconductivity 
\cite{Rokhsar1988} have also been recently subjected to similar investigations for ETH violation scenarios 
without disorder. Quantum many-body scars have been established to be a reason for the anomalous 
thermalization observed in a pedagogical model \cite{Bernien2017,Surace2020}: the Schwinger model with 
quantum link gauge fields \cite{Banerjee:2012pg}, also known as the PXP model \cite{Sachdev2002,Lesanovsky2012,Turner2018a} in 
condensed matter physics. Scars have been investigated in various kinds of Schwinger models and higher-spin 
PXP models \cite{Mukherjee2021spin1,Desaules:2022ibp,Desaules:2022kse}. Such anomalous states have also 
been uncovered in $\mathbb{Z}_2$ as well as certain non-Abelian gauge theories~\cite{Z2LGTscars, Hayata2023}.
It has been shown that even two-dimensional matter free $U(1)$ gauge theories possess a rich variety of scars 
\cite{Lan2017,Banerjee2021,Biswas:2022env}. In fact, the particular two-dimensional microscopic models are 
those which are used in the context of spin-ice (the quantum link model) \cite{Moessner2001,Shannon2004,Shannon2012}
and effective theories of quantum anti-ferromagnets (the quantum dimer model) 
 \cite{Rokhsar1988,Banerjee:2014wpa,Banerjee:2015pnt,Oakes_2018,Ran_2023}, and are known as Rokhsar-Kivelson 
models. In the latter case, the presence of exponentially many (in the longer linear dimension) quantum many-body 
scars could be shown for ladder systems, called lego scars~\cite{Biswas:2022env}. The structure of the quantum 
scars in the quantum link model has been comparatively elusive to identify. 

 In this article, we provide a systematic study of the quantum many-body scars of $U(1)$ quantum link models 
in their spin-$\frac{1}{2}$ representation and elucidate the structure of a class of scars. We call one 
variety of such scars as \emph{sublattice scars}. These scars have the peculiar property of being localized in 
one of the two possible sublattices of a $L_x \times L_y$ lattice with both $L_x, L_y$ being even due to
interference effects associated with the flipping of elementary plaquettes caused by the off-diagonal operator 
$\Okin$. While $\Okin$ has an exponentially large number (in $L_xL_y$) of exact zero modes due to a particular 
index theorem~\cite{Turner2018a,Turner2018b,Schecter2018index}, a typical zero mode is expected to mimic an 
infinite temperature thermal state locally and hence be completely featureless, as far as local features are 
concerned, from ETH. On the other hand, these sublattice scars display {\em perfect ordering} of the diagonal 
operator $\Opot$, which counts the total number of plaquettes which can be flipped. More specifically, the 
local operator $\Opots$ acting on each plaquette either takes its maximum value ($1$) or minimum ($0$) on the 
relevant sublattice. While most sublattice scars belong to the null space of $\Okin$, there also exist sublattice 
scars with nonzero integer eigenvalues of $\Okin$. We also unearth other anomalous zero modes of $\Okin$, or
quantum many-body scars, that are distinct from the sublattice scars. These scars turn out to be simultaneous 
zero modes of $\Okin$ and another non-commuting diagonal term composed by summing $\Opots$ over all elementary 
plaquettes of the lattice in a staggered fashion, depending on which sublattice the plaquette belongs to.

 The rest of the article is organized as follows: Sec \ref{sec:models} gives an account of two microscopic 
lattice gauge theory models with short-ranged Hamiltonians that we consider in our study, together with their 
local and global symmetries. Sec \ref{sec:scars} focuses on the description of sublattice scars that can 
either be anomalous zero modes of $\Okin$ or are eigenstates of $\Okin$ with eigenvalues $\pm 2$. We discuss 
a class of sublattice scars which can be analytically constructed using short singlets on a dual square lattice, 
as well as those which cannot be formed using such a description. In Sec \ref{sec:algo}, we describe an
efficient numerical method to specifically generate sublattice scars that goes beyond brute force exact 
diagonalization (ED). In Sec \ref{sec:parent}, we formulate a long-range {\em parent} Hamiltonian whose 
ground state consists of all the sublattice scars of the original short-range quantum link model with 
$\Okin=0$. The sublattice scars with eigenvalues $\Okin=\pm 2$ can be given a {\em quasiparticle-like} 
description starting from particular ground states of this parent Hamiltonian. Some interesting properties 
of the spectrum of this parent Hamiltonian are also discussed. In particular, zero energy states of this 
parent Hamiltonian turn out to be exact anomalous mid-spectrum zero modes of another short-ranged quantum 
$U(1)$ link  model, but with a staggered short-ranged diagonal term which is discussed in Sec.~\ref{sec:moreScars}. 
We finally conclude and present some open directions in Sec.~\ref{sec:conclusions}. 
 
\section{The models} \label{sec:models}
\begin{figure}[h]
    \includegraphics[scale=0.18]{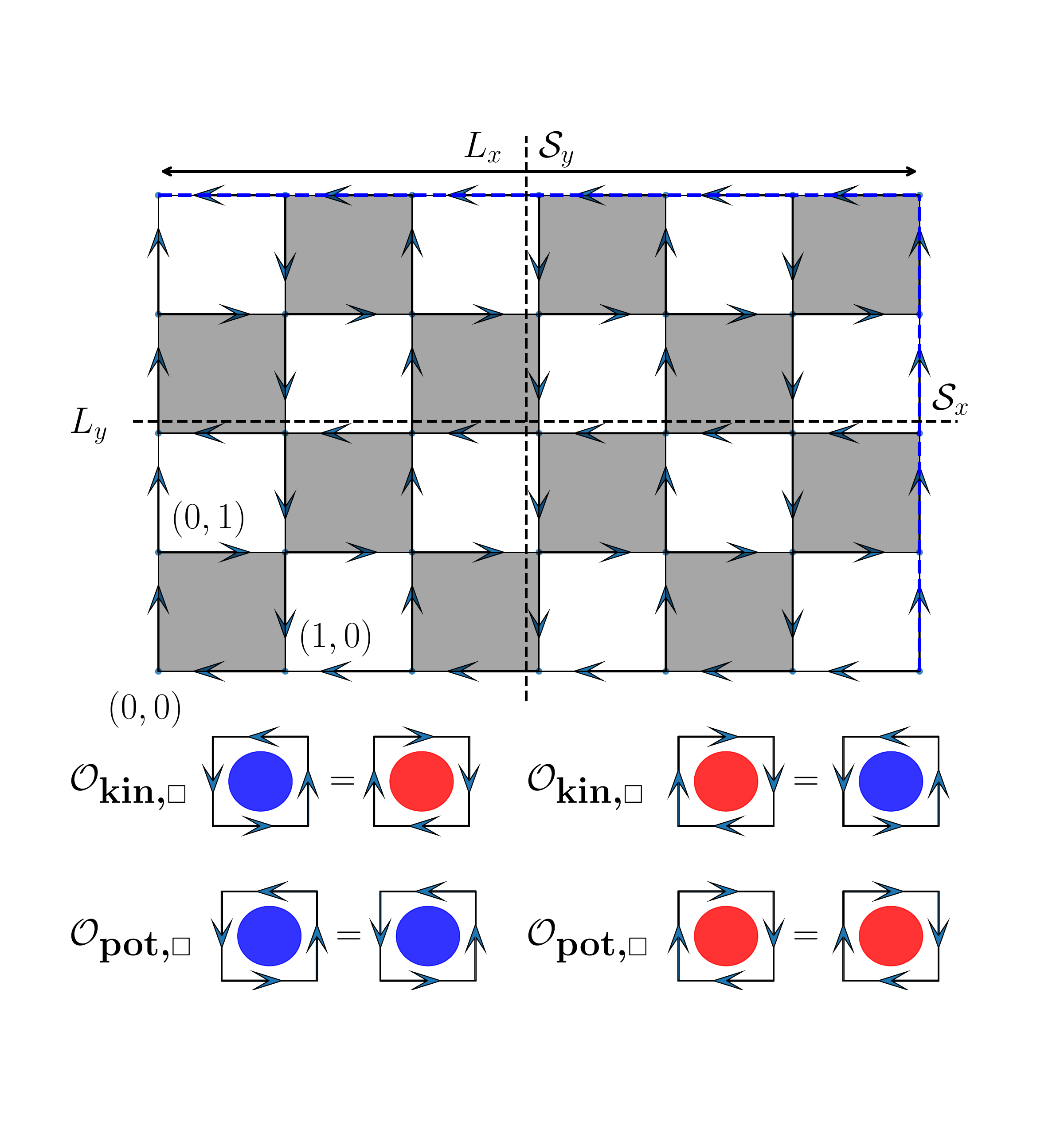}
    \caption{(Top panel) An electric flux configuration for a $(L_x,L_y)=(6,4)$ lattice with periodic 
    boundary condition in both directions. Even (odd) sub-lattice has been shown by shaded (blank) 
    plaquettes. $\mathcal{S}_x$ $(\mathcal{S}_y)$ is the reflection operation about $x$ ($y$)-axis 
    (shown by dotted lines). (Bottom panel) Action of $\Okins$ and $\Opots$ on elementary flippable 
    plaquettes. An elementary plaquette flippable in the clockwise (anti-clockwise) manner is shown 
    by red (blue) circle.}  
    \label{fig:qlmbasefig}
\end{figure}

 The investigation of ETH or violations thereof in the context of lattice gauge theories is complicated
by the fact that the traditional formulations of lattice gauge theories due to Wilson \cite{Kogut1975,Kogut1979}
use an infinite dimensional Hilbert space for each local degree of freedom. This allows for the presence of
arbitrarily high energy eigenstates in the spectrum. A better control of this ultraviolet divergence is provided
by the quantum link models \cite{Horn1981,Orland1989,Chandrasekharan1997} which still maintain exact gauge
invariance using finite dimensional gauge links. We continue using the smallest dimensional representation
for the $U(1)$ lattice gauge theory where the spin-$\frac{1}{2}$ operators are used for the gauge links 
as in the previous works \cite{Banerjee2021,Biswas:2022env}. The quantum links, as they are called, reside 
on the links ($r,\hat{\mu}$) connecting adjacent lattice sites $r$ and $r+\hat{\mu}$. The electric flux operator 
is given by, $E_{r,\hat{\mu}}=S^3_{r,\hat{\mu}}$ and the gauge fields are raising (lowering) operators, 
$U_{r,\hat{\mu}}(U^\dagger_{r,\hat{\mu}})=S^+_{r,\hat{\mu}}(S^-_{r,\hat{\mu}})$. The operator $\Okins$ is
composed of the elementary plaquette operator, 
$U_\square=U_{r,\hat{\mu}}U_{r+\hat{\mu},\hat{\nu}}U^\dagger_{r+\hat{\nu},\hat{\mu}}U^\dagger_{r,\hat{\nu}}$ 
and its hermitian conjugate $U^\dagger_\square$, which changes the direction of electric flux loops around an 
elementary plaquette (from clock-wise to anti clock-wise and vice versa) or annihilate non-flippable plaquettes.
The other operator we will use in order to construct microscopic Hamiltonians in the following subsection
is the operator $\Opots = \mathbb{P}_{r,\hat{\mu},\uparrow} \mathbb{P}_{r+\hat{\mu},\hat{\nu},\uparrow} 
 \mathbb{P}_{r+\hat{\nu},\hat{\mu},\downarrow}\mathbb{P}_{r,\hat{\nu},\downarrow} + \text{h.c}$. 
 $\mathbb{P}_{\uparrow (\downarrow)}$ is the projection operator to up (down) electric flux state of the
 gauge link. The structure
 of $\Opot$ counts the total number of flippable plaquettes: every flippable plaquette is counted as 1 
 irrespective of whether it is flippable in a clockwise or anti-clockwise manner, but as 0 if not flippable.
 A pictorial representation of the actions of $\Opots$ and $\Okins$ on a plaquette is illustrated in the 
 lower panel of Fig.~\ref{fig:qlmbasefig}.
 Note that the operators $\Opots$ and $\Okins$ use a four-body interaction, and the resulting theory is
 highly constrained: out of the 16 possible states at a single plaquette (for the spin-$\frac{1}{2}$
 representation) states, only two are non-trivially acted up by the operators $\Opots$ and $\Okins$. It is
 also useful to recognize that for the spin-$\frac{1}{2}$ representation $\Opots = \Okins^2$, with higher
 powers of $\Okins$ do not yield any new interactions which respect the gauge symmetry (which will be
 discussed next). As such in the lattice gauge theory set-up, this can be considered to be an adjoint
 interaction term for the gauge links. The other local interactions could be to take products of electric 
 fields on the four links of the plaquette, but oriented in different fashion, but it is unclear what such 
 interactions physically represent, and we do not consider such interactions in this work.

 The specific form of the Hamiltonians we build using the operators $\Opots, \Okins$ is explained in 
the next section, but before we discuss them, we consider the symmetries which are common to all the 
Hamiltonians considered in this article. The most relevant one is the local $U(1)$ symmetry, which 
arises due to the existence of the local operator, 
\begin{equation}
    G_r=\sum_\mu(E_{r,\hat{\mu}}-E_{r,r-\hat{\mu}})
    \label{eq:GL}
\end{equation}
which commutes with both the operators $[G_r, \Opot] = 0,~ [G_r, \Okin] = 0$, and thus with all Hamiltonians
which use these operators as building blocks. Moreover, note that the local operator is defined on the sites
and for the square lattice, and thus has four links touching each site. Because of this commutation relations,
the Hilbert space splits into exponentially many superselection sectors which then need to be specified by
imposing the Gauss' Law on particular states. We will give explicit examples when we discuss the model 
Hamiltonians. 

 Other symmetries of the Hamiltonians we will consider are all global. First, one has the discrete symmetries
such as lattice translations, lattice rotations, reflections, elements of the point group symmetries of the
square (rectangular) lattice. The presence or absence of some reflection symmetries will play a crucial 
role in our work. As an example, Fig.~\ref{fig:qlmbasefig} (Top panel), shows two particular reflection 
axes $\mathcal{S}_{x,y}$. Not all point-group symmetries commute with each other. Further, the Hamiltonians 
we consider are invariant to shifts by one lattice spacing, but one of the Hamiltonians we use will only be 
invariant by two lattice spacings. 
The charge conjugation $\mathbb{C}$ is another global
symmetry which transforms $^\mathbb{C} E_{r,\hat{\mu}} = - E_{r,\hat{\mu}}$, 
$^\mathbb{C} U_{r,\hat{\mu}} = U^\dagger_{r,\hat{\mu}}$,
and $^\mathbb{C} U^\dagger_{r,\hat{\mu}} = U_{r,\hat{\mu}}$. Finally, the model has a global 
$U(1) \times U(1)$ winding number symmetry corresponding to each spatial direction, generated by the operator 
$W_\mu = \frac{1}{L_\mu}\sum_r E_{r,\hat{\mu}}$. For a detailed discussion of the symmetries, the reader is 
referred to Ref.~\cite{Biswas:2022env}. We will, henceforth, focus on the largest sector with
zero winding, i.e., $(W_x,W_y)=(0,0)$.

\subsection{Two $U(1)$ quantum link lattice gauge theories}
 The first model we consider is the one corresponding to the $U(1)$ quantum link model, whose Hamiltonian is
given in terms of the plaquette operators introduced before, as follows:
\begin{equation}
\begin{split}
    \mathcal{H}_\mathrm{RK} & =\Okin + \lambda \Opot = -\sum_\square \Okins  + \lambda \sum_\square \Opots \\
    & =-\sum_\square (U_{\square} + U_{\square}^\dagger) 
    + \lambda \sum_\square (U_{\square} + U_{\square}^\dagger)^2.
\end{split}
\label{eq:H1}
\end{equation}
As noted previously, this Hamiltonian commutes with the Gauss Law operator in Eq.~\ref{eq:GL}. Different choices
of the Gauss Law decide different sectors, each with their own interesting physics. For example, demanding
that the vacuum is free of any charge implies the definition of physical states to be $G_r \ket{\psi} = 0$,
for all values of $r$.
The physics in this sector is relevant for the physics of confining Abelian gauge theories \cite{Banerjee2013},
as well as that of quantum spin-ice \cite{Shannon2004,Ran_2023}. Similarly, if one chooses the Gauss Law 
sector which consists of static (heavy) charges which are distributed throughout the lattice with the even-parity
sites having a charge $Q_r = +1$ and odd-parity sites having $Q_r = -1$, then the physics on the square lattice
is that of the quantum dimer model, relevant in the theories of high-temperature superconductivity
\cite{Rokhsar1988,Moessner2001,Banerjee:2015pnt}. These models have a rich phase diagram consisting of 
fractionally charged electric strings \cite{Banerjee:2014wpa} which break the translational symmetry and give
rise to crystalline confined phases, which confine static charges at zero temperature. At finite temperatures,
this model undergoes a phase transition into the deconfined phase, with the continuous phase transition
lying in the 2d XY universality class. Instead of single charges, if one uses charges $Q_r = \pm 2$ which hop 
on the lattice due to thermal fluctuations alone, then the resulting finite temperature phase transition
exhibits \emph{weak-universality}~\cite{Sau2023}.

 The second model we consider is very similar to the first Hamiltonian, except that the potential term carries
a staggered coupling:
\begin{equation}
\begin{split}
    \mathcal{H}_{\rm st} &  = -\sum_\square \Okins  + \lambda \sum_\square (-1)^\square \Opots \\
    & =-\sum_\square (U_{\square} + U_{\square}^\dagger) 
    + \lambda \sum_\square (-1)^{r_x+r_y} (U_{\square} + U_{\square}^\dagger)^2.
\end{split}
\label{eq:H2}
\end{equation}
 Note that each plaquette here is labelled by the lower left site, and therefore the sign 
 $(-1)^\square = (-1)^{r_x+r_y}$ is decided by whether the lower left site $r = (r_x,r_y)$ has even or odd parity. 

 We stress that while the Gauss Law, the winding numbers and the charge conjugation are the same for both the 
models, point group symmetries like reflections and translations act differently due to the presence/absence of 
the staggered coupling. E.g., while the first Hamiltonian is translationally invariant by one lattice spacing in both 
$\hat{\mu}=\hat{x},\hat{y}$, the second Hamiltonian is invariant to spatial translations by two lattice spacing 
due to the staggered couplings in $\mathcal{H}_{\mathrm{st}}$.

\subsection{An index theorem and mid-spectrum zero modes}
Both the models, $\mathcal{H}_{\mathrm{RK}}$ (Eq.~\ref{eq:H1}) and $\mathcal{H}_{\mathrm{st}}$ (Eq.~\ref{eq:H2}), 
are identical when $\lambda=0$ and describe a non-integrable $U(1)$ LGT~\cite{Banerjee2021,Biswas:2022env}. 
The spectrum of $\Okin$ has a spectral reflection symmetry due to the presence of an operator 
$\mathcal{C_\alpha}$ (where $\alpha=x,y$) that anticommutes with $\Okin$, where 
\begin{equation}
\mathcal{C}_\alpha  = \prod_{r,\alpha} E_{r,\alpha}.
\label{eq:chirality}
\end{equation}
Only the horizontal links with even (odd) $r_y$ [similarly, the vertical bonds with even (odd) $r_x$] 
contribute for $\alpha=x$ [similarly for $\alpha=y$] when $\frac{L_x}{2}$ is odd (even) [similarly, 
$\frac{L_y}{2}$ is odd (even)] in Eq.~\ref{eq:chirality}. For example, $\mathcal{C}_x$ is defined using 
the product of horizontal links on $\frac{L_y}{2}=2$ alternate rows with odd values of $r_y$ and  
$\mathcal{C}_y$ is defined using the product of vertical links on $\frac{L_x}{2}=3$ alternate columns 
with even values of $r_x$ for the 
$(6,4)$ lattice depicted in Fig.~\ref{fig:qlmbasefig}. This definition ensures that only one link in 
every elementary plaquette on the lattice contributes to $\mathcal{C}_\alpha$ from which it follows that 
\begin{equation}
\{\Okin,\mathcal{C}_\alpha\}=0.
\label{eq:index1}
\end{equation}
From the local constraint that $G_r \ket{\psi}=0$ at each lattice point $r$, we obtain that 
$\mathcal{C}_x \mathcal{C}_y=1$ implying that these are not independent of each other. For any eigenstate 
of $\Okin$ with energy $E \neq 0$, denoted by $\ket{E}$, there exists another eigenstate 
$\mathcal{C}_\alpha \ket{E}$ (where $\alpha$ may be chosen to be $x$ or $y$) with energy $-E$.

$\Okin$ commutes with the space reflection symmetries defined along the horizontal ($S_x$) or vertical 
($S_y$) axis (see Fig.~\ref{fig:qlmbasefig}) which divides the lattice in two equal halves leading to 
\begin{equation}
    [\Okin,S_\alpha]=0
    \label{eq:index2}
\end{equation}
for $\alpha=x,y$. Furthermore, in defining $\mathcal{C}_x$ and $\mathcal{C}_y$, the $\frac{L_y}{2}$ rows and 
the $\frac{L_x}{2}$ columns containing the links are located symmetrically with respect to the reflection axes 
$S_x$ and $S_y$ resulting in $\mathcal{C}_\alpha$ commuting with both $S_x$ and $S_y$, irrespective of 
$\alpha$, yielding 
\begin{equation}
    [\mathcal{C}_\alpha, S_\beta]=0
    \label{eq:index3}
\end{equation}
for $\alpha,\beta=x,y$. 

Remarkably, any Hamiltonian with these properties (Eq.~\ref{eq:index1}, Eq.~\ref{eq:index2}, 
Eq.~\ref{eq:index3}) has exact zero energy eigenstates with $E=0$ whose number scales exponentially with 
system size but which are, nonetheless, protected due to an index theorem as discussed in Refs.
~\cite{Turner2018a,Turner2018b,Schecter2018index}. Following these references, one can also show that the 
number of zero modes is bounded below by $\sqrt{\mathrm{HSD}}$ where $\mathrm{HSD}$ denotes the total 
Hilbert space dimension for a spin-$\frac{1}{2}$ QLM in a $(L_x,L_y)$ ladder in the appropriate winding 
number sector (zero winding number sector $(W_x,W_y)=(0,0)$ in our case). These zero modes of $\Okin$ 
turn out to be the only eigenstates that have a well-defined "chiral charge" of $\pm 1$ under the action 
of $\mathcal{C}_\alpha$ with $\alpha=x,y$ distinguishing them from all the non-zero modes of $\Okin$ 
which are not eigenstates of $\mathcal{C}_\alpha$. 

\begin{figure}[h!]
    \centering
    \includegraphics[scale=0.25]{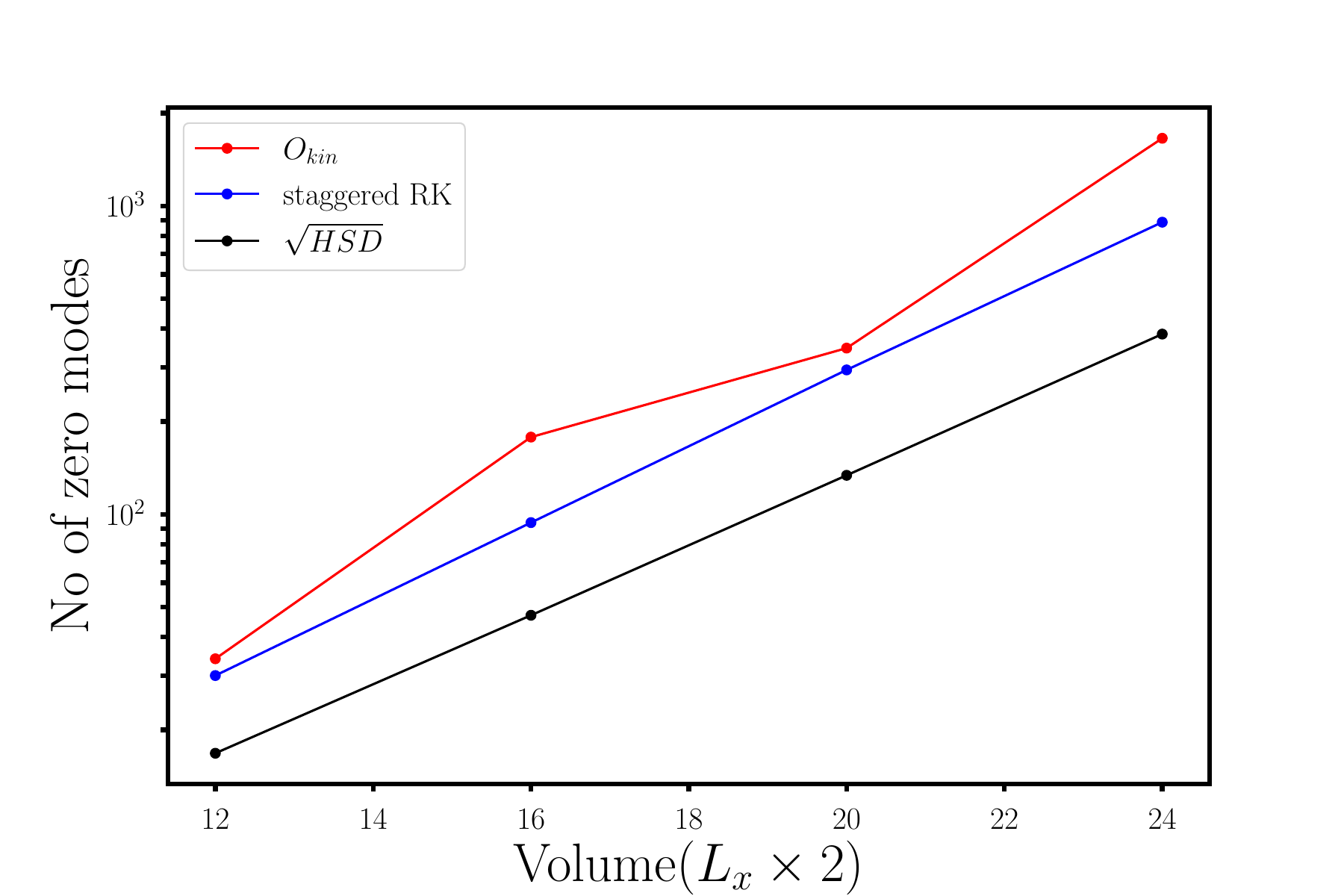}
    \caption{Scaling of the number of mid-spectrum zero modes with system size in two models which 
    satisfy index theorem. The behaviour of the lower bound $\sqrt{\mathrm{HSD}}$ is also shown.}
    \label{fig:Nzeroindex}
\end{figure}

The number of zero modes for $\Okin$ for a range of $(L_x,L_y)$ lattices that can be extracted from 
ED is shown in Table~\ref{tab:zromodes}. Table~\ref{tab:zromodes} also shows the Hilbert space dimension 
(HSD) for these lattices from exact enumeration.
\begin{table}[h!]
    \centering
    \begin{tabular}{|c|c|c|c|}
    \hline
      $L_x \times L_y$ & Zero modes in $\Okin$ & Zero modes in $\mathcal{H}_{\mathrm{st}}$ & HSD \\
    \hline
    $6\times2$ & 34 & 30 & 282 \\
    \hline
    $8\times2$ & 178 & 94 & 2214 \\
    \hline
    $10\times2$ & 346 & 294 & 17906 \\
    \hline
    $12\times2$ & 1658 & 886 & 147578 \\
    \hline
    $4\times4$ & 158 & 74 & 990 \\
    \hline
    $6\times4$ & 1070 & 426 & 32810 \\
    \hline
    \end{tabular}
    \caption{Number of mid-spectrum zero modes in $\Okin$ and $\mathcal{H}_{\mathrm{st}}$ and the 
    Hilbert space dimensions (HSD) for various lattices.}
    \label{tab:zromodes}
\end{table}

 It is useful to ask whether the index theorem is preserved when non-commuting terms like 
$\sum_{\square}\Opots$ or $\sum_{\square}(-1)^{\square}\Opots$ are added to $\Okin$. It turns 
out that adding $\sum_{\square}\Opots$ immediately violates the index theorem since an appropriate 
chiral operator which anticommutes with $\mathcal{H}_{\mathrm{RK}}$ cannot be constructed unless 
$\lambda=0$, making the spectrum lose its $E \rightarrow -E$ symmetry for $\lambda \neq 0$. On the 
other hand, adding the $\sum_{\square}(-1)^{\square}\Opots$ to $\Okin$ as done in 
$\mathcal{H}_{\mathrm{st}}$ presents an interesting case where the index theorem can be preserved 
for any $\lambda$. This is because while $\sum_{\square}(-1)^{\square}\Opots$ commutes with 
$\mathcal{C}_\alpha$ (for $\alpha=x,y$), it {\em anticommutes} with both $S_x$ and $S_y$ unlike 
$\sum_{\square}\Opots$ which {\em commutes} with $S_x$ and $S_y$. This allows for the weaker condition 
of $\{\mathcal{H}_{\mathrm{st}}, \mathcal{C}_\alpha S_\beta\}=0$ for $\alpha, \beta=x,y$ to be satisfied 
for any $\lambda$ which turns out to be sufficient to ensure the index theorem as demonstrated in 
Ref.~\cite{Turner2018b}. In fact, the number of zero modes of $\mathcal{H}_{\mathrm{st}}$ for a
$(L_x,L_y)$ lattice stays unchanged for any $\lambda \in (0, \infty)$ and is given in 
Table~\ref{tab:zromodes}. Note that it does not imply that the zero modes themselves stay unchanged 
as a function of $\lambda$.  Fig.~\ref{fig:Nzeroindex} shows the scaling of the number of zero modes 
of $\Okin$ and $\mathcal{H}_{\mathrm{st}}$ respectively and $\sqrt{\mathrm{HSD}}$ on various $(L_x,2)$ 
ladders to explicitly demonstrate the validity of the index theorem as well as the accuracy of the 
lower bound on the number of the zero modes. 

\section{Sublattice scars} \label{sec:scars}
Ref.~\cite{Banerjee2021} and Ref.~\cite{Biswas:2022env} discussed high-energy eigenstates 
of $\mathcal{H}_{\mathrm{RK}}$ (Eq.~\ref{eq:H1}) for $(L_x,L_y)$ lattices with periodic boundary conditions 
in both directions that stayed {\em unchanged} as a function of $\lambda$ by virtue of being simultaneous 
eigenstates of both $\Okin$ and $\Opot$. The fact that these high-energy states do not mix with the exponentially 
large (in system size) number of neighboring high-energy states as $\lambda$ is varied, immediately implies 
a violation of the ETH~\cite{Banerjee2021}. A class of these eigenstates have $\Opot=\frac{N_p}{2}$ (where 
$N_p=L_x L_y$ refers to the number of elementary plaquettes) and $\Okin=0$ or $\pm 2$ as ED studies of
Refs.~\cite{Banerjee2021,Biswas:2022env} showed. However, an analytic understanding of these scars in the 
QLM was lacking so far. We will give a physical description of these scars in what follows below, 
which will also demonstrate their existence in 2d lattices. 

It turns out that scars with $(\Okin,\Opot) = (0,\frac{N_p}{2})$ and $(\pm 2, \frac{N_p}{2})$ have much more structure 
than being just eigenstates of $\Opot$. {\em All} such scars for the finite lattices analyzed in 
Ref.~\cite{Biswas:2022env} can be understood from the following ansatz that the scar states $\ket{\psi_s}$ 
satisfy $\Opots \ket{\psi_s}=\ket{\psi_s}$ for one sublattice and $\Opots\ket{\psi_s}=0$ for the other 
sublattice (where there are two equivalent choices of which sublattice has $\Opots\ket{\psi_s}=\ket{\psi_s}$). 
Since these sublattice scars are mid-spectrum eigenstates of a non-integrable theory~\cite{Biswas:2022env}, 
namely $\Okin$, ETH would have predicted $\Opots\ket{\psi_s}=\frac{1}{2}\ket{\psi_s}$, independent of which 
sublattice an elementary plaquette is located. Thus, these sublattice scars clearly violate ETH. We will 
discuss a class of sublattice scars that have a description in terms of emergent short singlets on an 
appropriately defined dual square lattice in Sec.~\ref{sec:singlets} and will also place a lower bound 
on the number of these scars for arbitrary (even) $(L_x,L_y)$ lattices. Not all sublattice 
scars have such an analytic description and these non-singlet scars will be discussed in 
Sec.~\ref{sec:nonsinglets} and Sec.~\ref{sec:nonzeroOkin}. The understanding of the structure of these 
non-singlet scars was greatly helped by an efficient numerical algorithm to specifically target 
sublattice scars (but not the full Hilbert space) and would be discussed in the next section.

\subsection{Short singlet sublattice scars}
\label{sec:singlets}
 A class of sublattice scars with $\Okin=0$ can be constructed exactly using a tiling representation 
in terms of emergent singlets. Let us denote the plaquettes with $\Opots=1$ ($\Opots=0$) as active 
(inactive). For sublattice scars, any active plaquette is surrounded by four inactive 
plaquettes as its nearest neighbors (sharing a common link) and vice versa. Now, consider 
two nearest neighbor active plaquettes which share a common vertex. These plaquettes can only have 
the following four local configurations, $\left(\fbox{C},\fbox{A}\right)$, 
$\left(\fbox{A},\fbox{C}\right)$, $\left(\fbox{C},\fbox{C}\right)$, and  
$\left(\fbox{A},\fbox{A}\right)$, where $\fbox{C}$ $\left(\fbox{A}\right)$ denotes an elementary 
plaquette where the electric fluxes have a clockwise (anti-clockwise) circulation.
The inactive plaquettes on the other sublattice are denoted by $\fbox{U}$ henceforth. It is important 
to stress here that $\Opots=0$ on a plaquette does not fix its electric flux configuration and allows 
for electric flux fluctuations since there are $14$ local flux configurations that are $\fbox{U}$ on 
any elementary plaquette. 
\begin{figure}[h!]
    \centering
    \includegraphics[scale=0.15]{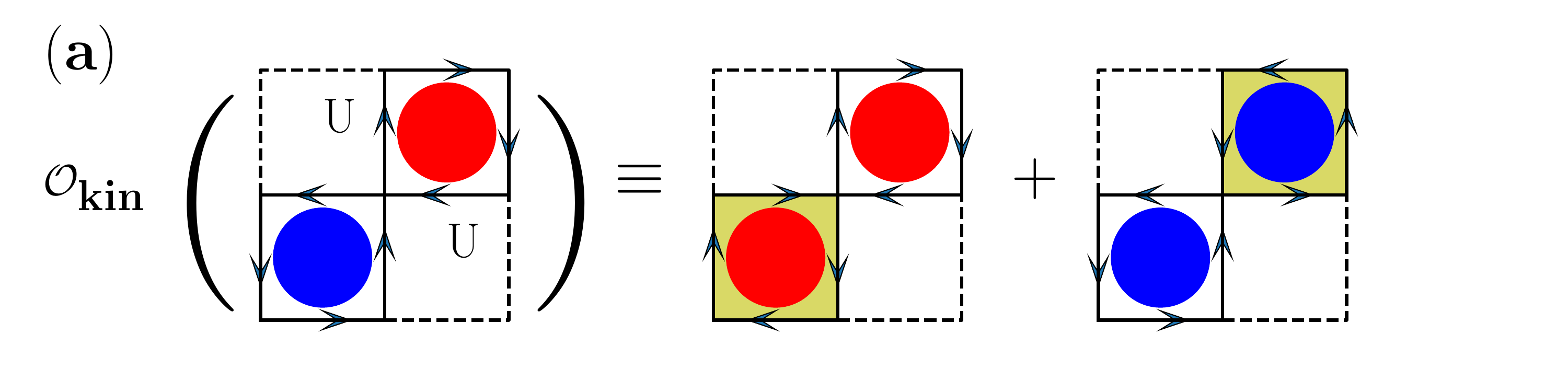}
    \includegraphics[scale=0.15]{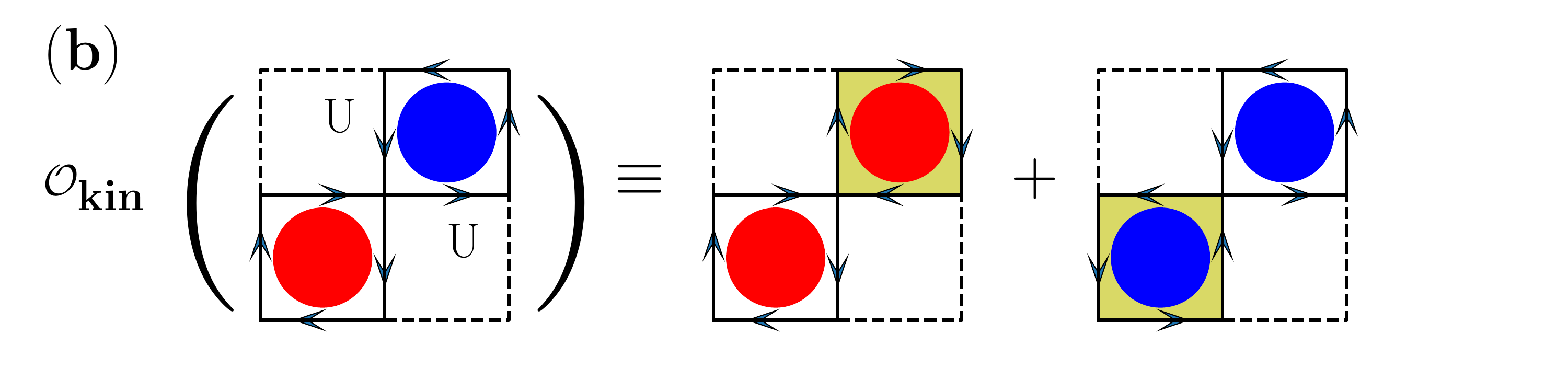}
    \includegraphics[scale=0.15]{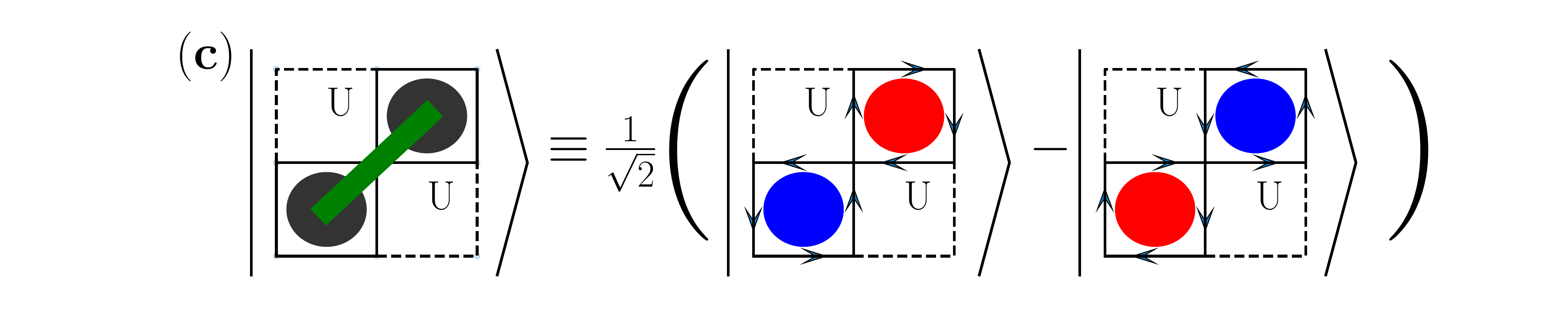}
    \caption{(a) and (b) show the action of $\Okin$ on two $2 \times 2$ units composed of two 
    unflippable plaquettes and one clockwise and one anti-clockwise flippable plaquette. Clockwise 
    (anti-clockwise) circulation of electric fluxes on a plaquette is shown by a red (blue) circle 
    as in Fig.~\ref{fig:qlmbasefig}. The plaquette marked by a yellow background indicates the one 
    on which $\Okin$ acts in panels (a) and (b). (c) Representation of an emergent singlet in the 
    same $2 \times 2$ unit.}
    \label{fig:cancellation}
\end{figure}

Let us first consider two nearest neighbor active plaquettes that are either in the local configuration 
$\left(\fbox{A},\fbox{C}\right)$ or $\left(\fbox{C},\fbox{A}\right)$. This automatically ensures that 
the two plaquettes on the other sublattice that share edges with both these active plaquettes are 
$\fbox{U}$ without the need of specifying the electric fluxes on any other links of these plaquettes. 
The above statement is not true if the local configuration of two nearest neighbor active plaquettes 
is either  $\left(\fbox{A},\fbox{A}\right)$ or $\left(\fbox{C},\fbox{C}\right)$ and enforces extra 
constraints on the state of the other active plaquettes that are in contact with the two $\fbox{U}$ 
plaquettes to ensure their unflippability. We can now take this $2\times 2$ unit of two flippable 
and two unflippable plaquettes and apply $\Okin$ on this unit. This gives a configuration as shown 
in Fig.~\ref{fig:cancellation} (a), (b) from which it is clear that acting $\Okin$ on a {\em singlet} 
of $\frac{1}{\sqrt{2}}\left(\fbox{A},\fbox{C}-\fbox{C},\fbox{A} \right)$ annihilates the state 
(Fig.~\ref{fig:cancellation}(c)).   

\begin{figure}[h!]
    \centering
    \includegraphics[scale=0.22]{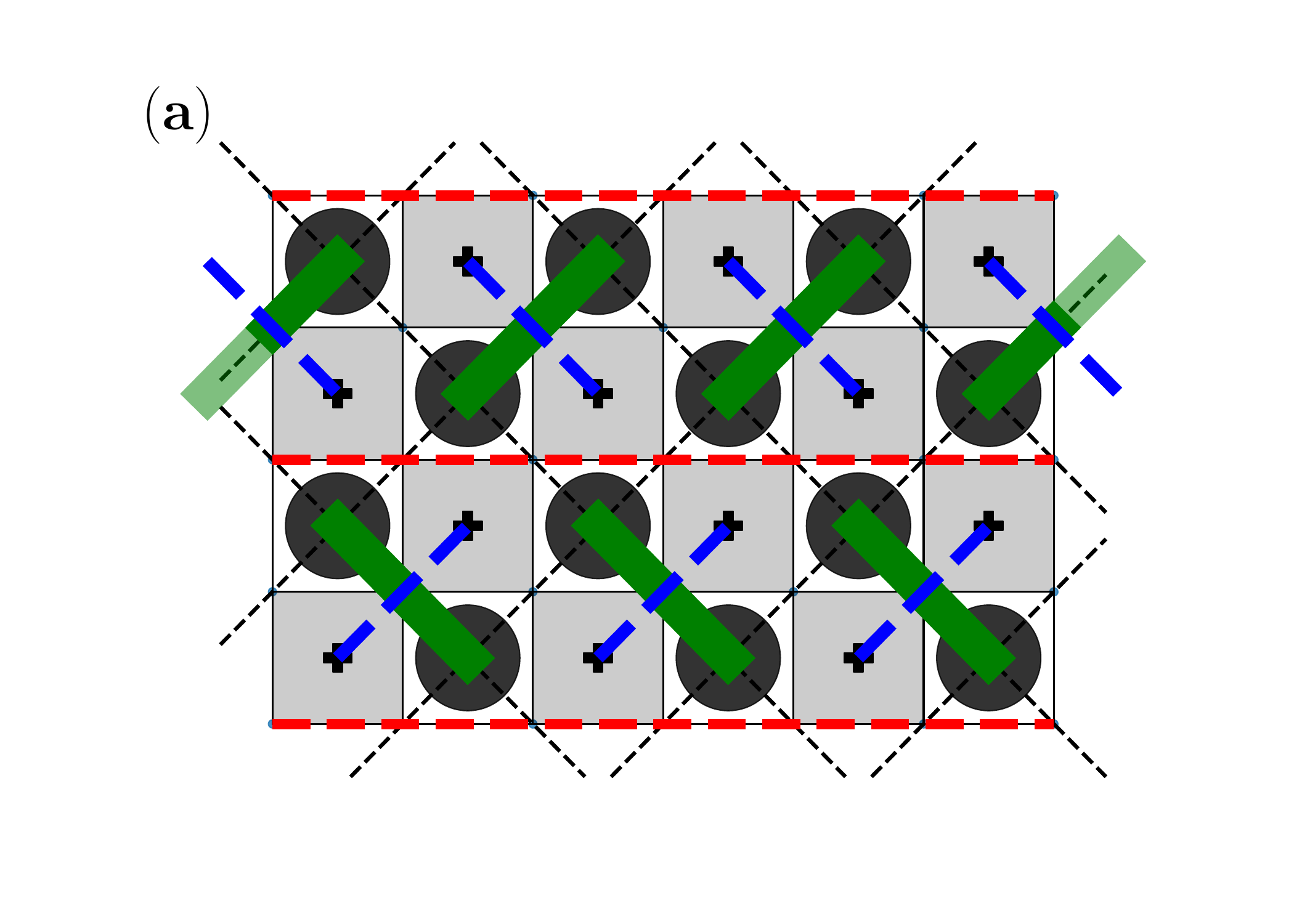}
    \includegraphics[scale=0.12]{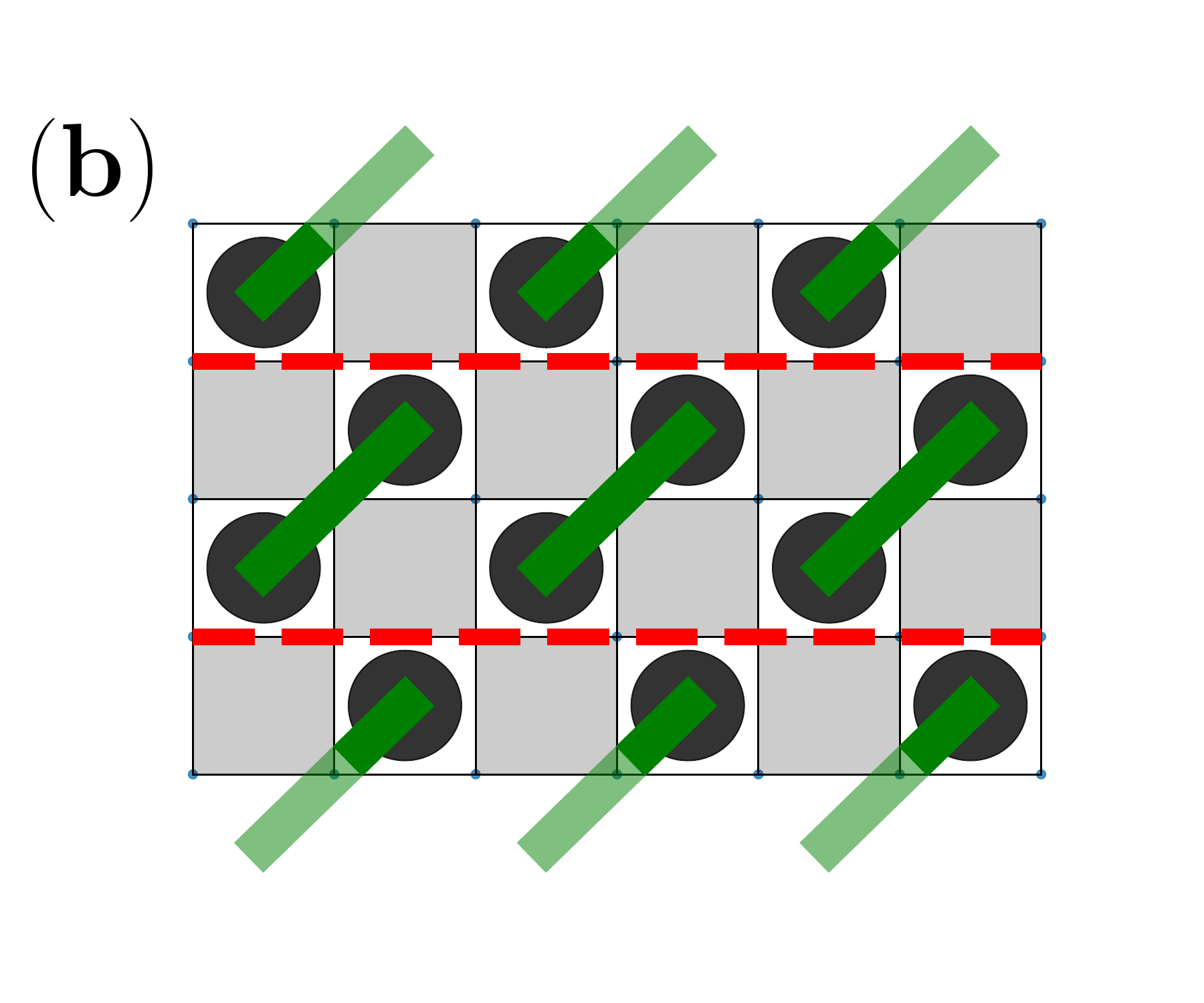}
    \includegraphics[scale=0.12]{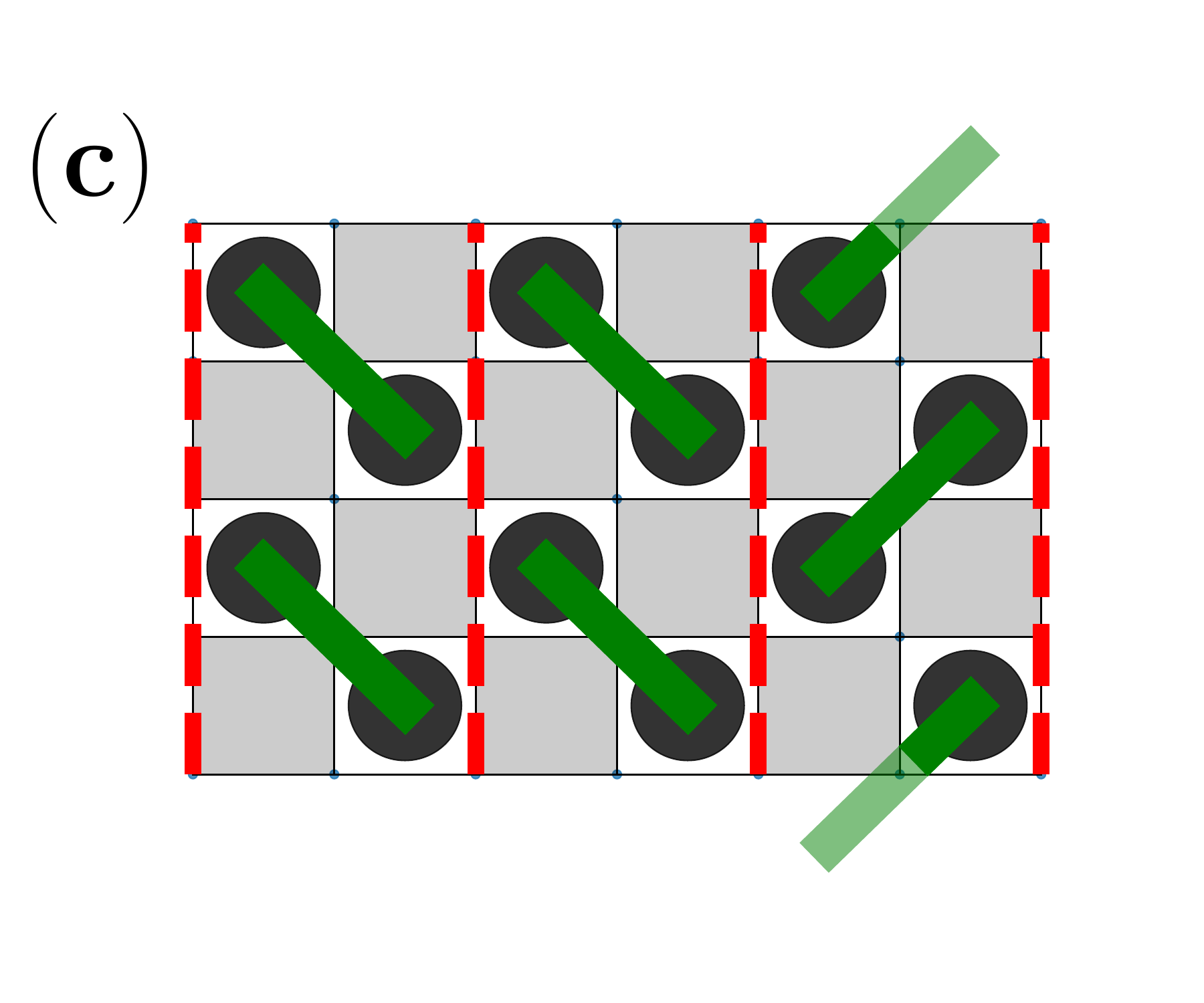}
    \caption{Graphical representations of some short singlet sublattice scars for a $(6,4)$ lattice. 
    The white (grey) plaquettes in all panels denote active (inactive) plaquettes with $\Opots=1 (0)$.
    The singlets (or dimers), shown as thick green lines, follow the same convention as shown in 
    Fig.~\ref{fig:cancellation}(c) and reside on the bonds of the dual square lattice formed by connecting 
    the centers of the active plaquettes as indicated in panel (a) by thin dotted lines. Each dimer 
    has an associated perpendicular bisector as shown in panel (a) by dotted blue lines. Panels (a) 
    and (b) show two horizontal partitions of width $2$ while panel (c) shows $3$ vertical partitions 
    of width $2$ using dotted red lines.}
    \label{fig:shortsinglets}
\end{figure}

We now ask whether these $2 \times 2$ units comprising a singlet and two unflippable plaquettes 
can be used to tile the entire $(L_x,L_y)$ lattice and give sublattice scars. We define
a {\em dual} square lattice by joining the centers of the active plaquettes on one sublattice
(Fig.~\ref{fig:shortsinglets}(a)). The emergent singlets can be thought of as living on the bonds 
of this dual lattice. These singlets naturally act as hard-core dimers because any active plaquette 
cannot be part of more than one singlet. Furthermore, the perpendicular bisector of each 
such dimer touches the center of the two unflippable plaquettes in the $2 \times 2$ unit 
(Fig.~\ref{fig:shortsinglets}(a)). Since the number of unflippable plaquettes equals $\frac{N_p}{2}$, none 
of these perpendicular bisectors associated with the hard-core dimers (representing the singlets) 
can share ends and thus act as {\em dual} dimers with their own hard-core repulsion on the square 
lattice defined by bonds touching the centers of the inactive plaquettes. The allowed 
coverings of these dimers (i.e., singlets) and their associated perpendicular bisectors (the dual dimers) 
define all sublattice scars with a simple emergent singlet description. Each of these short singlet 
sublattice scars are eigenstates of $\Okin$ with eigenvalue $0$ and satisfy 
$\Opots \ket{\psi}= c \ket{\psi}$, where $c$ = $1$ ($0$) for the active (inactive) elementary plaquettes. 

 Let us now discuss the degeneracy of such singlet sublattice scars on $(L_x,L_y)$ lattices. Consider
the $(L_x,2)$ ladder first, and focus on active plaquettes on one particular sublattice. It can be seen 
that the hard-core constraints on the singlets and their associated perpendicular bisectors only allow 
for two possible coverings such that all short singlets are parallel to each other with their possible 
alignment providing the additional degeneracy of $2$. Since the choice of active plaquettes on one 
sublattice is arbitrary, the total number of such short singlet sublattice scars equals $2+2=4$ for 
$(L_x,2)$ ladders.   

We next consider the degeneracy of short singlet sublattice scars for wider ladders with $(L_x,4)$. 
The valid dimer coverings can be constructed by first dividing the lattice into a close packing of 
parallel non-overlapping horizontal or vertical partitions (see Fig.~\ref{fig:shortsinglets} for 
some examples), each of width $2$, and then arranging parallel dimers in each partition in one of 
the two possible orientations. The internal orientation of the parallel dimers can be assigned in 
each of the partitions independently. For $(L_x,4)$ lattices, there are two ways to divide the 
lattice into $\frac{4}{2}=2$ ($\frac{L_x}{2}$) non-overlapping horizontal (vertical) partitions taking into 
account periodic boundary conditions. This gives the total number of dimer coverings to be 
$\left(8+2^{\frac{L_x}{2}+1}\right)$ out of which $4$ dimer coverings that comprise all dimers 
being parallel to each other are repeated by multiple partitions. Thus, the total number of 
{\em distinct} dimer coverings on one particular sublattice of active plaquettes equals 
$\left(8+2^{\frac{L_x}{2}+1}-4\right)$. Not all these dimer coverings are linearly independent states. 
An explicit calculation shows that For a $(4,4)$ lattice, $10$ out of $12$ such dimer coverings are 
linearly independent scars while the number appears to be  $\left(3+2^{\frac{L_x}{2}+1}\right)$ for 
$L_x>4$ (see Table.~\ref{table:scarOkin0}). Extending this counting to an arbitrary $(L_x,L_y)$ lattice, 
we thus obtain $\mathcal{O}(2^{\frac{L_x}{2}+1}+2^{\frac{L_y}{2}+1}-4)$ such short singlet sublattice 
scars which immediately provides a lower bound on the number of sublattice scars.

 Let us briefly discuss the symmetries of these scars. While the short singlet sublattice scars do not, 
in general, have a well-defined momentum since the dimer covering may not have any particular periodicity, 
all these states are eigenstates of the charge conjugation $\mathbb{C}$. Since the emergent singlet is 
odd under $\mathbb{C}$, a short singlet scar composed of even (odd) number of singlets has 
$\mathbb{C}=+1$ ($-1$). Since one needs $\frac{L_x L_y}{4}$ short singlets to form a sublattice scar, 
these particular scars have $\mathbb{C}=(-1)^{\frac{L_x L_y}{4}}$ for a $(L_x,L_y)$ lattice.  

\subsection{Non-singlet sublattice scars}
\label{sec:nonsinglets}
	\begin{table}[h]
	\begin{tabular}{|p{1.5cm}|l|l|l|}
		\hline
		\multirow{2}{*}{Lattice}
		& \multicolumn{2}{|c|}{\parbox{2.5cm}{Scars in equal A-C sector}} & \multicolumn{1}{|c|}{\parbox{2cm}{Scars in unequal A-C sector}} \\
		\cline{2-3}
		& Singlet scars & {{Non-singlet scars}} &  \\ 
		\hline
      $L_x\times2$ & 2 & 0 & 0 \\ 
		\hline
		 $4\times4$ & 10 & 0 & 3 \\ 
		\hline
		 $6\times4$ & 19 & 3 & 1 \\ 
		 \hline
		 $8\times4$ & 35 & 17 & 1 \\ 
		\hline
	   $10\times4$ & 67 & 62 & 1 \\ 
		\hline
      $6\times6$ & 28 & 1 & 1 \\ 
		\hline
	\end{tabular}
	\caption{Number of sublattice scars with $\Okin=0$ in one sublattice for various lattices. 
    The sublattice scars can be further classified into short singlet scars, non-singlet scars formed out of 
    Fock states with an equal (unequal) number of clockwise (represented by C) and anti-clockwise (represented 
    by A) flippable active plaquettes. The corresponding degeneracies have been separately listed for clarity. }
    \label{table:scarOkin0}
	\end{table}

From our numerical algorithm that explicitly targets sublattice scars (see Sec.~\ref{sec:algo} for a 
discussion of the procedure), we can compute the total number of sublattice scars in finite $(L_x,L_y)$ 
lattices. This information is given in Table~\ref{table:scarOkin0} for sublattice scars with $\Okin=0$ 
where the active plaquettes (with $\Opots=1$) have been chosen to be on one particular sublattice. While 
all sublattice scars for $L_y=2$ turn out to be short singlet scars, wider lattices with $L_y \geq 4$ 
immediately lead to the presence of non-singlet sublattice scars. From explicit calculations, we see 
that sublattice scars with $\Okin=0$ can be written as a linear combination of Fock states that involve 
either $\left(\frac{N_p}{4},\frac{N_p}{4} \right)$ $\left(\fbox{A}, \fbox{C} \right)$ or 
$\left(\frac{N_p}{4} \pm 1, \frac{N_p}{4} \mp 1 \right)$ $\left(\fbox{A}, \fbox{C} \right)$ active 
plaquettes (Table.~\ref{table:scarOkin0}). Clearly, short singlet scars belong to the former variety 
while the latter variety cannot have a short singlet representation. Focusing on sublattice scars that 
can be expressed as a linear combination of Fock states with $\left(\frac{N_p}{4},\frac{N_p}{4} \right)$ 
$\left(\fbox{A}, \fbox{C}\right)$ active plaquettes, we see that while $(4,4)$ lattice does not have 
any non-singlet scar, $L_x \geq 6$ for $L_y=4$ has several non-singlet scars whose number seems to 
rapidly increase with $L_x$. It is not clear to us whether this number in fact exceeds the corresponding 
number of singlet scars when $L_x>10$ for $L_y=4$. For $(6,6)$ there is exactly $1$ such non-singlet scar 
but we expect that this number again increases as $L_x>6$ for $L_y=6$. Note that accessing even $(10,4)$ 
lattices with ED targeting the full Hilbert space is extremely resource-intensive and here we have used 
the algorithm presented in Sec.~\ref{sec:algo}. 

\begin{figure}[h!]
           \centering
           \includegraphics[width=.55\textwidth,right]{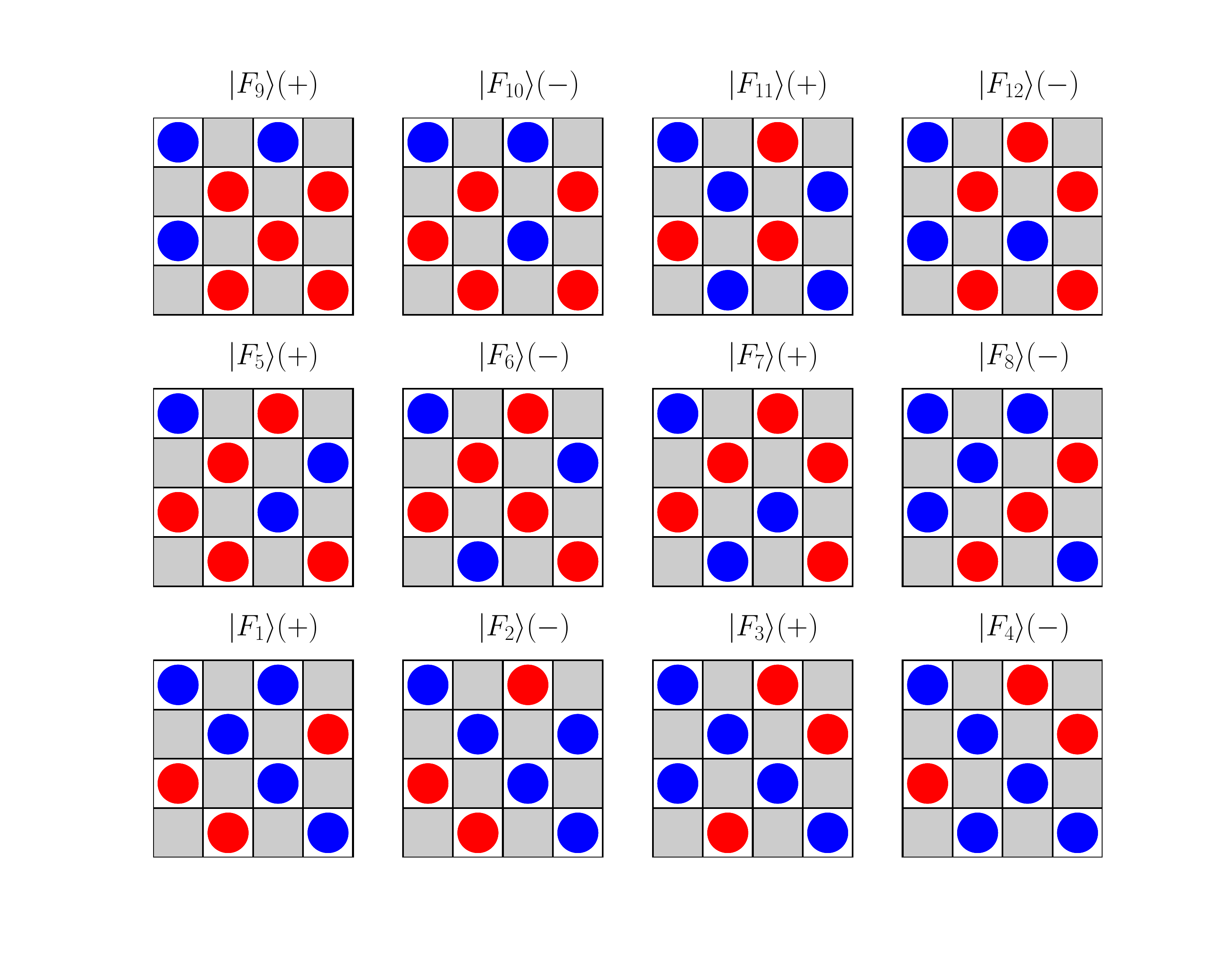}
           \caption{One of the three sublattice scars with $\Okin=0$ and 
           having contributions from unequal number of clockwise and anti-clockwise 
           flippable plaquettes for $4\times4$ lattice. Clockwise (anti-clockwise) circulation 
           of electric fluxes on a plaquette is shown by a red (blue) circle as in Fig.~\ref{fig:qlmbasefig}. 
           The white (grey) plaquettes denote active (inactive) plaquettes with $\Opots=1 (0)$ 
           as in Fig.~\ref{fig:shortsinglets}.}
           \label{fig:confuneq4x4}
       \end{figure}

We now come to the sublattice scars that can be expressed in terms of Fock states with 
$(\frac{N_p}{4}+1,\frac{N_p}{4}-1)$ and $(\frac{N_p}{4}-1,\frac{N_p}{4}+1)$ 
$\left(\fbox{A}, \fbox{C}\right)$ active plaquettes on one particular sublattice. Interestingly, their 
number does not seem to increase with lattice dimension unlike the singlet and non-singlet scars composed 
of an equal number of clockwise and anti-clockwise active plaquettes (Table.~\ref{table:scarOkin0}). While 
there are $3$ such sublattice scars for a $(4,4)$ lattice, there is only $1$ such scar for $(6,4)$, $(8,4)$, 
$(10,4)$ and $(6,6)$ lattices. A particular linear combination of the $3$ sublattice scars (which is also 
a sublattice scar, by definition) for a $(4,4)$ lattice is shown in Fig.~\ref{fig:confuneq4x4}. This scar 
has the property that all the Fock states that combine to form this state contribute with equal
magnitudes but with an intricate sign structure as depicted in Fig.~\ref{fig:confuneq4x4}, and the state 
can be expressed as
\begin{equation}
  \ket{\psi_{s,0}} = \frac{1}{2\sqrt{6}} \sum_{i=1}^{12} \mathrm{Sign}(i)\left(\ket{F_i} 
  + \mathbb{C} \ket{F_i} \right),
  \label{eq:unequalstate4times4}
\end{equation}
where we refer the reader to Fig.~\ref{fig:confuneq4x4} for $\mathrm{Sign}(i)$ for the corresponding 
$\ket{F_i}$. This striking property of the amplitudes of the contributing Fock states turns out to be 
true for the unique scar with $(\frac{N_p}{4}+1,\frac{N_p}{4}-1)$ and $(\frac{N_p}{4}-1,\frac{N_p}{4}+1)$ 
$\left(\fbox{A}, \fbox{C}\right)$ active plaquettes on one particular sublattice for bigger lattices 
than $(4,4)$.  

 We comment on the symmetries of the non-singlet scars. We observe that the
eigenvalues of the operator $\mathbb{C}$ for these states are $(-1)^{\frac{L_x L_y}{4}}$, just 
like the short singlet scars, though we do not have an analytic proof in this case. Secondly, 
the unique scars formed by combining Fock states with $(\frac{N_p}{4}+1,\frac{N_p}{4}-1)$ and 
$(\frac{N_p}{4}-1,\frac{N_p}{4}+1)$ $\left(\fbox{A}, \fbox{C}\right)$ active plaquettes on one 
particular sublattice for $(6,4), (8,4), (10,4)$ and $(6,6)$ lattices have a well-defined momentum 
with respect to translations by two lattice units in both directions $\hat{x}$ and $\hat{y}$ 
(the sublattice structure rules out any symmetry of these states with respect to translations by 
one lattice unit). We denote the momentum as $\hat{k}_x, \hat{k}_y$ to avoid any confusion
with the usual momentum, defined through single lattice translations.
While these scars for $((6,4), (8,4), (10,4)$ lattices have a momentum 
$(\hat{k}_x,\hat{k}_y)=(0,0)$, the scar for the $(6,6)$ lattice is at momentum 
$(\hat{k}_x,\hat{k}_y)=(\pi,\pi)$. The unique non-singlet scar for a $(6,6)$ lattice 
that involves equal number of clockwise and anti-clockwise plaquettes on one sublattice also 
carries a momentum $(\hat{k}_x,\hat{k}_y)=(\pi,\pi)$. 

\begin{figure}
    \centering
    \includegraphics[scale=0.25]{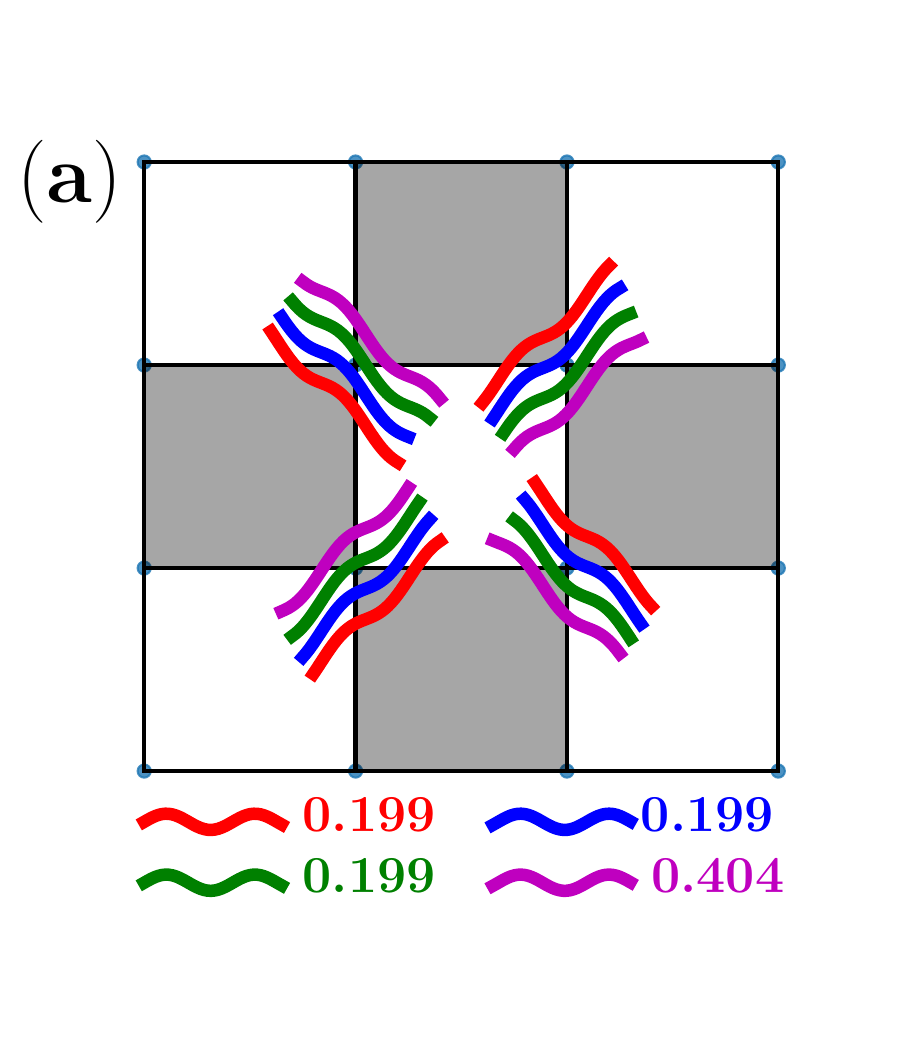}
    \includegraphics[scale=0.25]{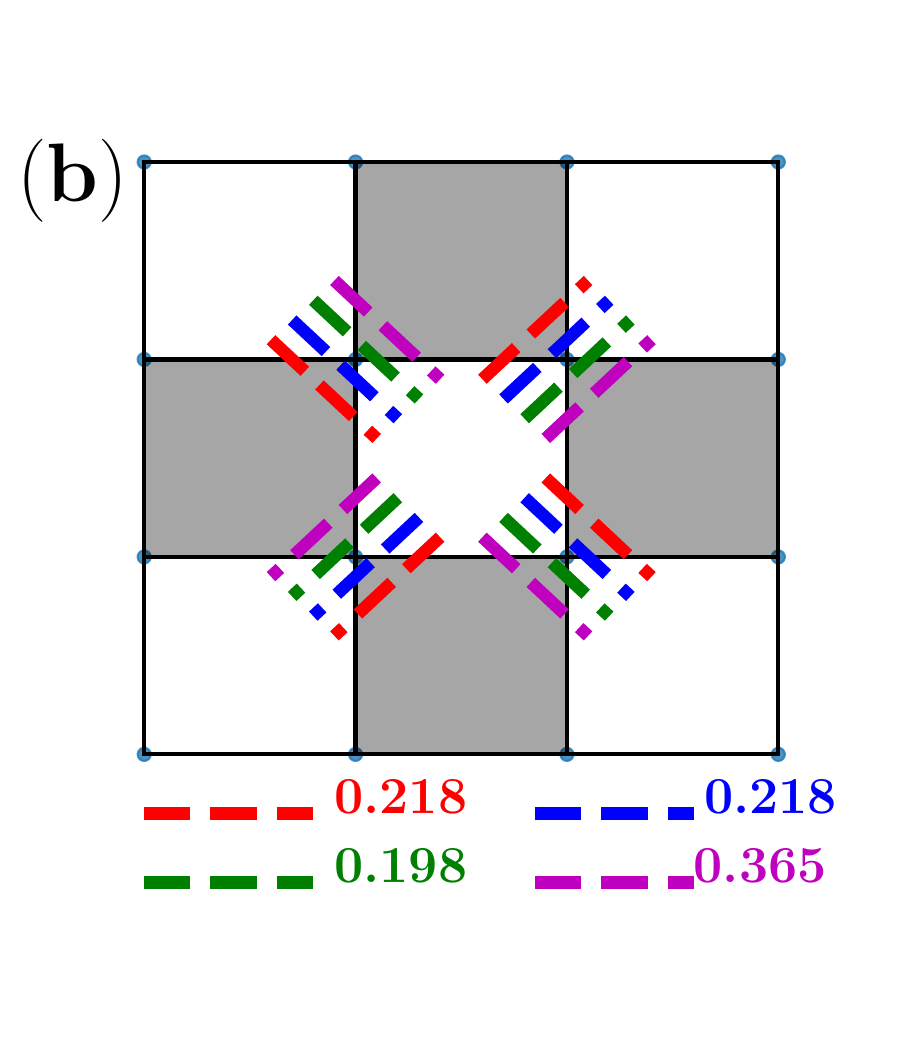}
    \caption{Expectation value $\braket{t_{+1}}_{(r_1,r_2)}$ (red lines), $\braket{t_{-1}}_{(r_1,r_2)}$ 
    (blue lines), $\braket{t_{0}}_{(r_1,r_2)}$ (green lines) and $\braket{s_{0}}_{r_1,r_2}$ (magenta lines) 
    for the unique non-singlet scars on a $(6,6)$ lattice composed of (a) equal number of clockwise and 
    anticlockwise active plaquettes, and (b) unequal number of clockwise and anticlockwise active plaquettes. 
    The white (grey) plaquettes denote active (inactive) plaquettes with $\Opots=1 (0)$ as in 
    Fig.~\ref{fig:shortsinglets}.}
    \label{fig:correlation}
\end{figure}

 One can ask how the local operators differ in any non-singlet scar compared to any short singlet scar. 
These operators must be other than $\Opots$ which cannot distinguish between the two cases. 
However, certain two-plaquette correlation functions, where the two plaquettes are nearest 
neighbor active plaquettes sharing a common vertex, are sensitive to the scar states. 
We expect that these operators can be measured straightforwardly in quantum simulator experiments,
and provide a route to the experimental demonstration of these states.
The four-dimensional local Hilbert space can then be represented in terms of \emph{singlets} and 
\emph{triplets} in the following manner:
       \begin{eqnarray}
         \ket{t_{+1}} &=& \left(\fbox{A}, \fbox{A} \right) \nonumber \\
         \ket{t_{-1}} &=& \left(\fbox{C}, \fbox{C} \right) \nonumber \\
         \ket{t_{0}}  &=& \frac{1}{\sqrt{2}}\left(\fbox{A}, \fbox{C} + \fbox{C}, \fbox{A}\right) \nonumber \\
         \ket{s_{0}}  &=& \frac{1}{\sqrt{2}}\left(\fbox{A}, \fbox{C} - \fbox{C}, \fbox{A}\right)
         \label{eq:twoplaqoperator}
       \end{eqnarray}
 We can then probe the expectation values of the operators $\ket{t_{\pm 1,0}} \bra{t_{\pm 1,0}}$ 
and $\ket{s_0} \bra{s_0}$ locally for all nearest neighbor active plaquettes given any sublattice scar, 
which we denote by a shorthand $\braket{t_{\pm 1,0}}_{(r_1,r_2)}$, $\braket{s_0}_{(r_1,r_2)}$ where 
$(r1,r_2)$ indicates the bond connecting the centers of the two plaquettes. For a short singlet 
sublattice scar, $\braket{s_0}_{(r_1,r_2)} = 1$ and $\braket{t_{\pm 1,0}}_{(r_1,r_2)} = 0$ for 
the bonds that carry a singlet (dimer) (Fig.~\ref{fig:shortsinglets}) and 
$\braket{t_{\pm 1,0}}_{(r_1,r_2)} = \braket{s_0}_{r_1,r_2} = \frac{1}{4}$ for the other bonds 
(Fig.~\ref{fig:shortsinglets}). For non-singlet sublattice scars, we choose to probe these local 
operators for both the unique non-singlet sublattice scar with equal number of clockwise and 
anticlockwise active plaquettes (Fig.~\ref{fig:correlation}(a)) and with 
$(\frac{N_p}{4} \pm 1, \frac{N_p}{4} \mp 1)$ clockwise and anticlockwise active plaquettes 
(Fig.~\ref{fig:correlation}(b)) on a $(6,6)$ lattice. For both these non-singlet scars, the 
values of $\braket{t_{\pm 1,0}}_{(r_1,r_2)}$, and $\braket{s_0}_{(r_1,r_2)}$ are independent 
of the location of the bond $(r_1,r_2)$ and are quite different from short singlet scars. 
Probing these operators locally for the unique non-singlet scar with $(\frac{N_p}{4} \pm 1, N_p \mp 1)$ 
clockwise and anticlockwise active plaquettes for $(L_x,4)$ lattices with $L_x \geq 6$ again gives 
values of $\braket{t_{\pm 1,0}}_{(r_1,r_2)}$, and $\braket{s_0}_{(r_1,r_2)}$ that are independent 
of the location of the bond $(r_1,r_2)$. We obtain $\braket{t_{\pm 1}}_{(r_1,r_2)}=\frac{1}{4}$ 
for $L_x=6, 8, 10$ respectively, $\braket{t_{0}}_{(r_1,r_2)} (\braket{s_{0}}_{(r_1,r_2)})\approx 
0.2038 (0.29167)$ for $L_x=6$, $\braket{t_{0}}_{(r_1,r_2)} (\braket{s_{0}}_{(r_1,r_2)}) \approx 
0.2187 (0.2812)$ for $L_x=8$ and $\braket{t_{0}}_{(r_1,r_2)} (\braket{s_{0}}_{(r_1,r_2)}) 
\approx 0.225 (0.275)$ for $L_x=10$.

\subsection{Sublattice scars with $\Okin=\pm 2$}
\label{sec:nonzeroOkin}
\begin{table}[h]
   	\begin{tabular}{|c|c|c|}
   		\hline
   		Lattice & Scars with $\Okin=+2$ & Scars with $\Okin=-2$ \\
   		\hline
   		$4\times4$ & 3 & 3\\
   		\hline
   		$6\times4$ & 1 & 1\\
   		\hline
   		$8\times4$ & 1 & 1\\
   		\hline
   		$6\times6$ & 1 & 1\\
   		\hline
   	\end{tabular}
    \caption{Number of sublattice scars with $\Okin=\pm 2$ for various lattice dimensions.}
    \label{table:scarOkin2}
   \end{table}
      
 We also find sublattice scars with $\Okin = \pm 2$ for systems with $L_y \geq 4$ 
(see Table.~\ref{table:scarOkin2}) that are simultaneous eigenstates of $\Opots$ 
with it being equal to $1$ on one sublattice and $0$ on another. Since these states 
have non-zero integer values of $\Okin$, these necessarily violate the ETH as a generic 
high-energy state would instead have an irrational energy eigenvalue. It is 
interesting to ask how such simple eigenvalues may be generated from $\Okin$ since 
there is no analogous index theorem that ensures the presence of specific $E \neq 0$ 
eigenstates with such simple energies. Comparing Table~\ref{table:scarOkin0} and Table~\ref{table:scarOkin2}, 
it is striking that the degeneracy of sublattice scars with $\Okin=+2 (-2)$ equals 
that of sublattice scars with $\Okin=0$ that are composed of Fock states with unequal 
number of clockwise and anticlockwise active plaquettes.

 We now demonstrate a ``triangle relation'' between sublattice scars with $\Okin = \pm 2$ 
and the sublattice scars with $\Okin=0$, composed of Fock states with $(\frac{N_p}{4} \pm 1, \frac{N_p}{4} \mp 1)$ 
clockwise and anticlockwise active plaquettes. For this, we note that the sublattice scars 
with $\Okin=+2 (-2)$ consist of both Fock states with $(\frac{N_p}{4}, \frac{N_p}{4})$ and 
$(\frac{N_p}{4} \pm 1, \frac{N_p}{4} \mp 1)$ clockwise and anticlockwise active plaquettes. 
Given any $\ket{\psi_{s,+2}}$ (where $+2$ denotes the $\Okin=+2$ states), applying the chirality 
operator $\mathcal{C}_\alpha \ket{\psi_{s,+2}}$ generates a unique $\ket{\psi_{s,-2}}$ with 
$\Okin=-2$ ($\alpha$ can be chosen to be $x$ or $y$). Applying $\mathcal{C}_\alpha$ 
on the Fock states with $(\frac{N_p}{4}, \frac{N_p}{4})$ [$(\frac{N_p}{4} \pm 1, \frac{N_p}{4} \mp 1)$] 
clockwise and anticlockwise active plaquettes gives an eigenvalue of +1 [-1]. 
Thus, $\left(\ket{\psi_{s,+2}} - \mathcal{C}_\alpha \ket{\psi_{s,+2}} \right)$ leads to a state 
with only Fock states with $(\frac{N_p}{4} \pm 1, \frac{N_p}{4} \mp 1)$ clockwise and anticlockwise 
active plaquettes contributing to it. These Fock states turn out to be \emph{identical} to the ones 
that make up the sublattice scars with $\Okin=0$, and unequal number of clockwise and anticlockwise 
active plaquettes, but the sign structure of the states are nonetheless different.

 We now define an operator $\mathcal{O}$ that acts on Fock states with $(\frac{N_p}{4} \pm 1, 
 \frac{N_p}{4} \mp 1)$ clockwise and anticlockwise active plaquettes and leads to a sign change 
 (no sign change) for Fock states with $(\frac{N_p}{4} + 1, \frac{N_p}{4} - 1)$ 
 ($(\frac{N_p}{4} - 1, \frac{N_p}{4} + 1)$) clockwise and anti-clockwise active plaquettes. We then 
 see that
\begin{equation}
  \mathcal{O}\left(\ket{\psi_{s,+2}} - \mathcal{C}_\alpha \ket{\psi_{s,+2}} \right) \propto \ket{\psi_{s,0}}
  \label{eq:trianglerelation}
  \end{equation}
where $\ket{\psi_{s,0}}$ is a sublattice scar with $\Okin=0$ composed of Fock states with unequal 
number of clockwise and anticlockwise active plaquettes on one sublattice. We note that this relation 
is also true for $(4,4)$ system where there is a three-fold degeneracy of sublattice scars with 
$\Okin=+2 (-2)$. We conjecture that such a ``triangle relation'' exists for all $(L_x,L_y)$ lattices 
with $L_y \geq 4$.

 Before closing this section, we note that sublattice scars with $\Okin=+2 (-2)$ are eigenstates of 
charge conjugation with $\mathbb{C}=-1$ for $(L_x,L_y)$ lattices with $L_y=4$ and with $\mathbb{C}=+1$ 
for $(6,6)$ lattice. Additionally, the unique sublattice scars with $\Okin=+2 (-2)$ for $(6,4)$, $(8,4)$, 
and $(6,6)$ lattices have a well-defined momentum of $(\hat{k}_x,\hat{k}_y)=(0,0)$ with respect to 
translations by two lattice units on both directions.

\section{Efficient algorithm to generate sublattice scars} 
\label{sec:algo}

 In this section, we discuss the efficient numerical procedure to specifically target
the sublattice scars, unlike exact diagonalization (ED), which constructs all states full Hilbert 
space. While both algorithms are computationally exponentially expensive, the former diverges slower,
as can be guessed from Tab.~\ref{tab:zromodes}, and will be demonstrated further in Tab.~\ref{tab:SVDtable}. 
For this, we start with a set of $n$ Fock states 
$\{ \ket{f_i}\}_n$ that have the requisite sublattice structure, i.e., all plaquettes on one (other) sublattice 
are flippable (non-flippable). We will soon come to the question of how to generate this set efficiently. Our 
task is to now find states of the form  $\ket{\psi_s}=\sum_{i=1}^n a_i \ket{f_i}$ such that 
$\Okin \ket{\psi_s} = \mathcal{N} \ket{\psi_s}$, where $\mathcal{N}$ is either zero or a non-zero integer. 
Given $\{\ket{f_i}\}_n$, we define a set of $m$ Fock states $\{\ket{F_i}\}_m$ which contains all possible 
states generated by a single action of $\Okin$ on the set $\{\ket{f_i}\}_n$. This set can be written as 
$\{\ket{F_i}\}_m = \{\ket{f_i}\}_n + \{\ket{f_i^\prime}\}_{m^\prime}$, where  $\{\ket{f_i^\prime}\}_{m^\prime}$ 
contains Fock states which \emph{do not} have the sublattice structure of all flippable (non-flippable) 
plaquettes on one (other) sublattice. One can write 
$\Okin \ket{f_i} = \sum_{j=1}^m V_{ji} \ket{F_j} = \sum_{j=1}^n v_{ji} \ket{f_j} 
+ \sum_{j=1}^{m^\prime} v_{ji}^\prime \ket{f_j^\prime}$ where $V_{ji}=\braket{F_j|\Okin|f_i},\;
v_{ji}=\braket{f_j | \Okin | f_i}$ and $v_{ji}^\prime = \braket{f_j^\prime | \Okin |f_i}$.

First, we consider the sublattice scars with $\Okin \ket{\psi_s}=0$. We then have 
$\Okin \ket{\psi_s} = \sum_{i=1}^{n}a_i \sum_{j=1}^m V_{ji} \ket{F_j} = 
\sum_{j=1}^{m} (\sum_{i=1}^{n} V_{ji}a_i) \ket{F_j}$. To satisfy $\Okin \ket{\psi_s}=0$, 
we need to satisfy $m$ simultaneous conditions: $\sum_{i=1}^{n} V_{ji}a_i=0$ for $j=1$ to $m$. 
The possible set of solutions ($\{a_i\}$) specify the sublattice scars with $\Okin=0$. For this, 
we construct the $m \times n$ dimensional matrix $[\mathbb{V}_{m\times n}]_{ji}=v_{ji}$. If we 
perform a singular value decomposition (SVD) of $\mathbb{V}$, then the right-singular vectors 
($\{[\Phi_{n\times1}]\}$) with zero-singular value satisfy the above conditions because 
$[\mathbb{V}_{m\times n}][\Phi_{n\times1}]=[\O_{m\times1}] \Rightarrow 
\sum_{i=1}^{n} V_{ji}[\Phi_{n\times1}]_{i1}=0$ for $j=1$ to $m$. These right-singular vectors 
are then the required sublattice scars with $\mathcal{N}=0$. 

 Next, we consider the sublattice scars with $\Okin \ket{\psi_s} = \mathcal{N} \ket{\psi_s}$ where 
$\mathcal{N}$ is a non-zero integer. This requires a more complicated procedure. We can write 
$ \Okin \ket{\psi_s } = \sum_{j=1}^n (\sum_{i=1}^n v_{ji}a_i) \ket{f_j} 
+ \sum_{j=1}^{m^\prime} (\sum_{i=1}^n v_{ji}^\prime a_i) \ket{f_j^\prime}$. To satisfy 
$ \Okin \ket{\psi_s} = \mathcal{N} \ket{\psi_s}$ with $\mathcal{N} \neq 0$, we need to 
satisfy two conditions simultaneously: (i) $\sum_{i=1}^n v_{ji}a_i=\mathcal{N}a_j$ for $j=1$ to $n$ 
and (ii) $\sum_{i=1}^n v_{ji}^\prime a_i=0$ for $j=1$ to $m^\prime$. First, we construct the matrix 
$[\mathbb{V}^\prime_{m^\prime\times n}]_{ji}=v_{ji}^\prime$. As we have already argued before, the 
right-singular vectors of $\mathbb{V}^\prime$ with zero-singular value satisfy condition (ii). Let 
us denote the space spanned by these right-singular vectors with zero singular value (assume that 
there are $p$ of them) $\{ \ket{\Psi_1}, \ket{\Psi_2}, \cdots, \ket{\Psi_p}\}$ by $\mathbb{Z}_V$. 
The required sublattice scars are those which reside in $\mathbb{Z}_V$ and satisfy condition (i). 
We first construct the matrix $[\Tilde{\mathbb{V}}_{n\times n}]_{ji}=(v_{ji}-\mathcal{N}\delta_{ji})$ 
and then take the projection to $\mathbb{Z}_V$: 
$[(\widetilde{\mathbb{V}}_0)_{n\times p}]=[\Tilde{\mathbb{V}}_{n\times n}]\times[\mathbb{M}_{n\times p}]$. 
The columns of $\mathbb{M}$ are formed by the column matrices $\ket{\Psi_i}$s. Again, the right-singular 
vectors of $\widetilde{\mathbb{V}}_0$ with zero singular value will satisfy both conditions (i) and (ii). 
Using this method, we could only find solutions for $\mathcal{N} = \pm 2$ for the system sizes that we 
could numerically access. 

 We can decompose the sublattice scars into Fock states with equal (unequal) number of clockwise and 
anti-clockwise flippable plaquettes on one sublattice. To do that, we simply divide the set 
$\{\ket{f_i}\}_n$ into the sets of equal and unequal clockwise and anticlockwise flippable plaquettes, 
$\{\ket{f_i^{\mathrm{eq}}}\}_{n_{\mathrm{eq}}}$ and $\{\ket{f_i^{\mathrm{uneq}}}\}_{n_{\mathrm{uneq}}}$. 
Then using the above method for the individual sets, we can find if any scar can be obtained from 
$\{\ket{f_i^{\mathrm{eq}}}\}_{n_{\mathrm{eq}}}$ and $\{\ket{f_i^{\mathrm{uneq}}}\}_{n_{\mathrm{uneq}}}$. 
At this point, it is worth stressing that the above method of obtaining sublattice scars is more efficient 
than the full ED, because in this method we use a small subspace of the full Hilbert space (see 
Tab.~\ref{tab:SVDtable} for a comparison). 

\begin{table}[h!]
    \centering
    \begin{tabular}{|c|c|c|c|c|c|}
        \hline
        Lattice & HSD & $n_{\mathrm{eq}}$ & $m_{\mathrm{eq}}$ & $n_{\mathrm{uneq}}$ & $m_{\mathrm{uneq}}$ \\
        \hline
        $6\times4$ & 32810 & 510 & 1392 & 552 & 1628\\
        \hline
        $8\times4$ & 1159166 & 4662 & 16352 & 5984 & 20574\\
        \hline
        $10\times4$ & 42240738 & 43896 & 186040 & 61720 & 248302\\
        \hline
        $6\times6$ & 5482716 & 13778 & 55080 & 19260 & 72818\\
        \hline
    \end{tabular}
    \caption{Hilbert space dimension (HSD) and number of Fock states present in 
    $\{\ket{f_i^{\mathrm{eq}}}\}_{n_{\mathrm{eq}}}$, $\{\ket{F_i^{\mathrm{eq}}}\}_{m_{\mathrm{eq}}}$, 
    $\{\ket{f_i^{\mathrm{uneq}}}\}_{n_{\mathrm{uneq}}}$ and $\{\ket{F_i^{\mathrm{uneq}}}\}_{m_{\mathrm{uneq}}}$ 
    for various lattices.}
    \label{tab:SVDtable}
\end{table}

\begin{figure}[h!]
    \centering
    \includegraphics[width=0.18\textwidth]{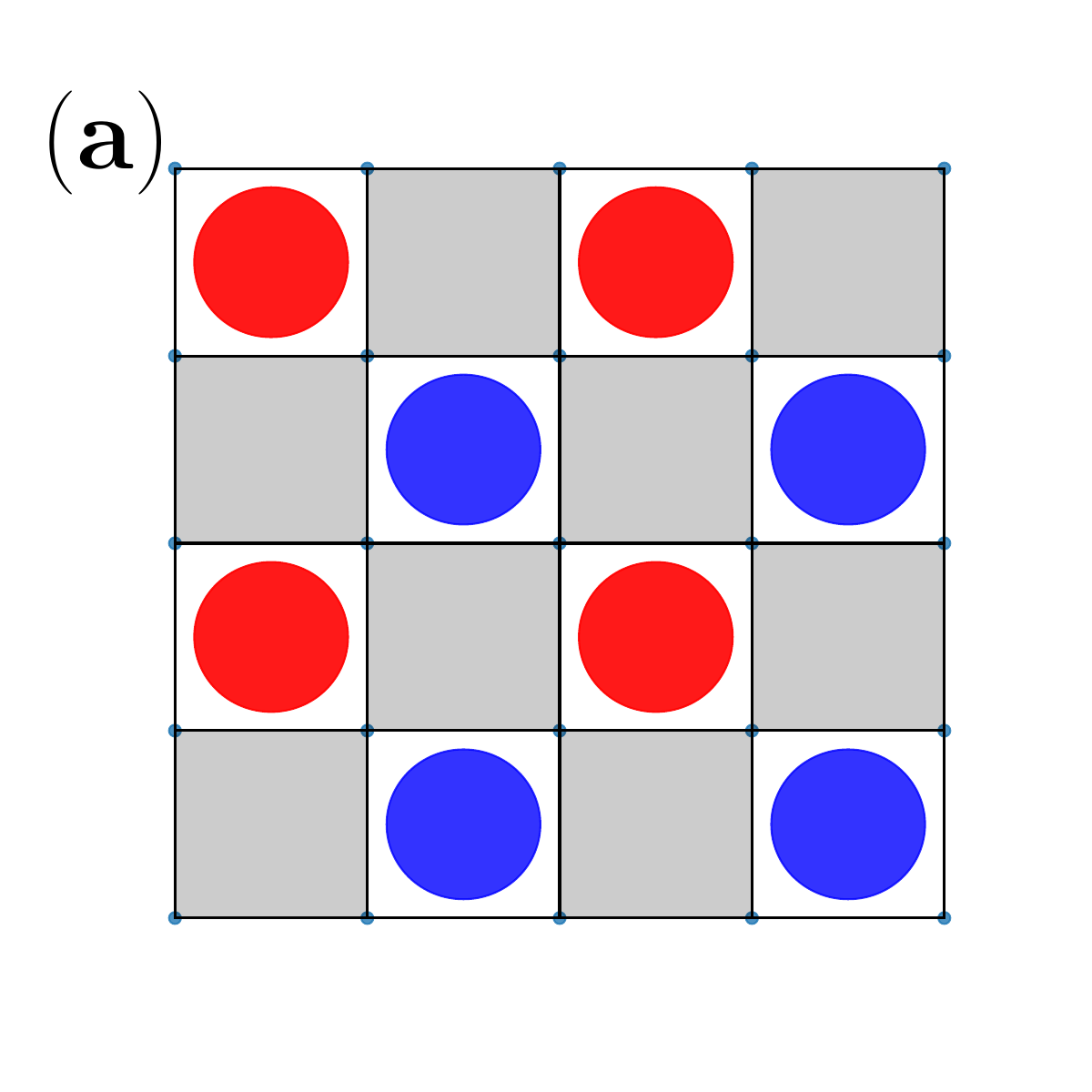}
    \includegraphics[width=0.18\textwidth]{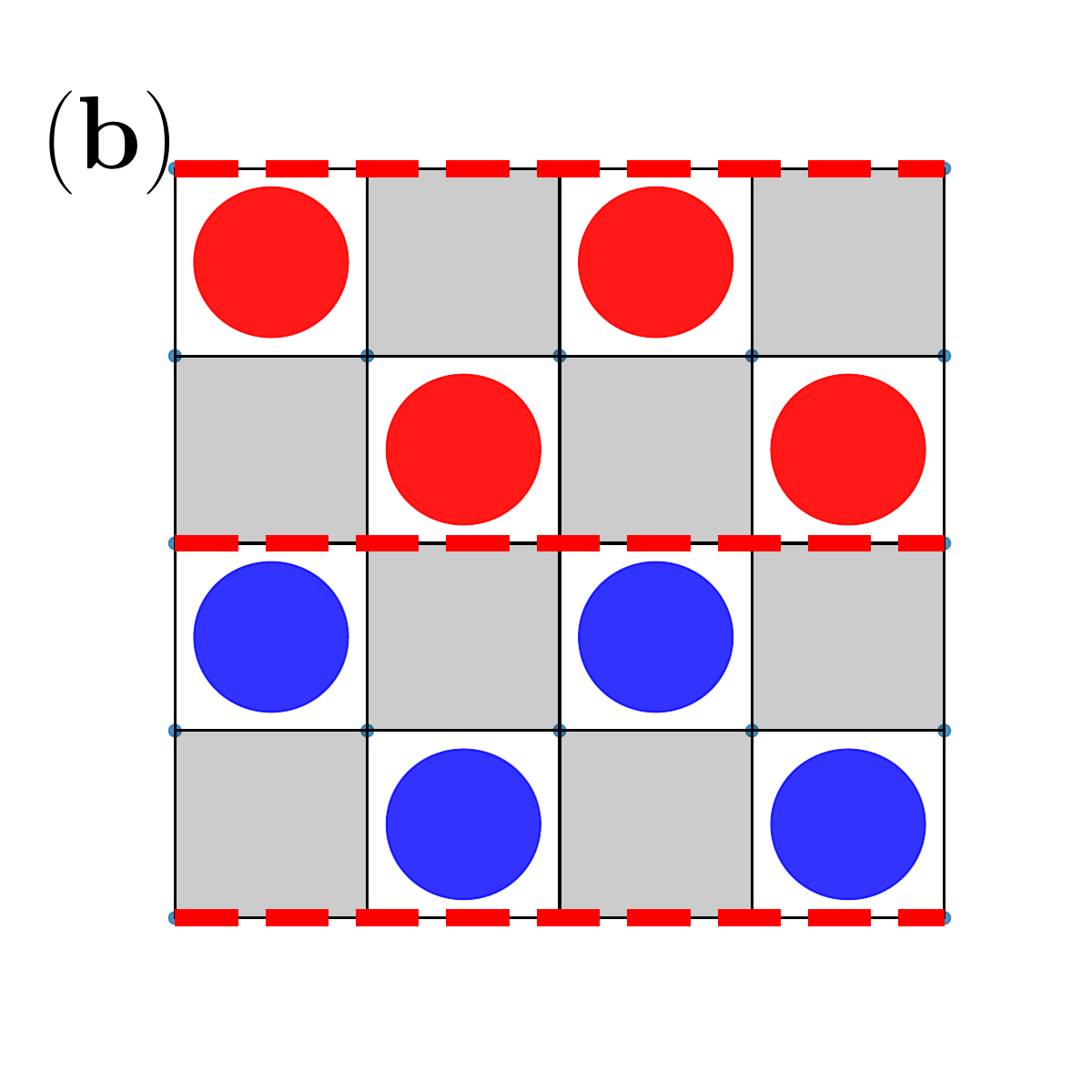}
    \includegraphics[width=0.18\textwidth]{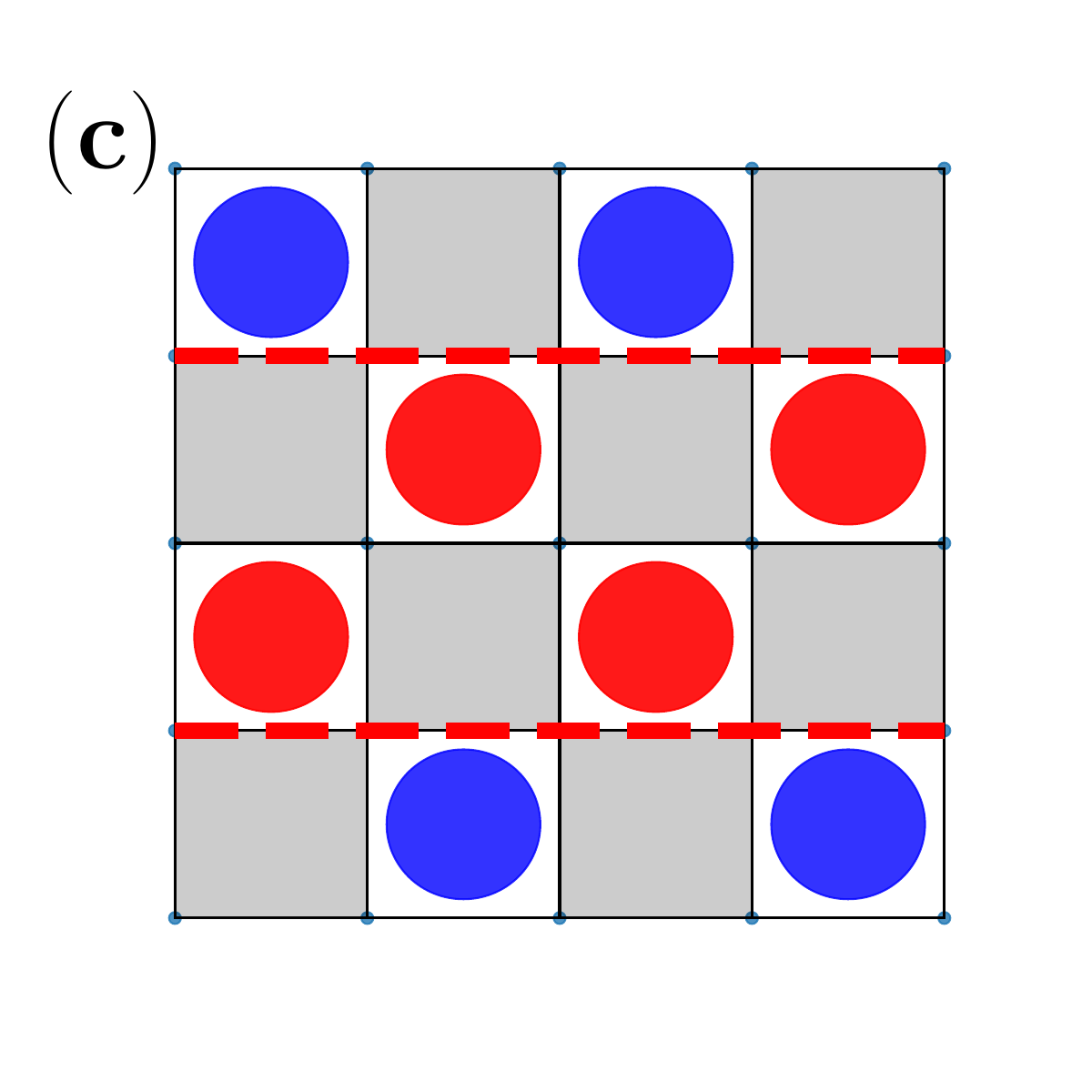}
    \includegraphics[width=0.18\textwidth]{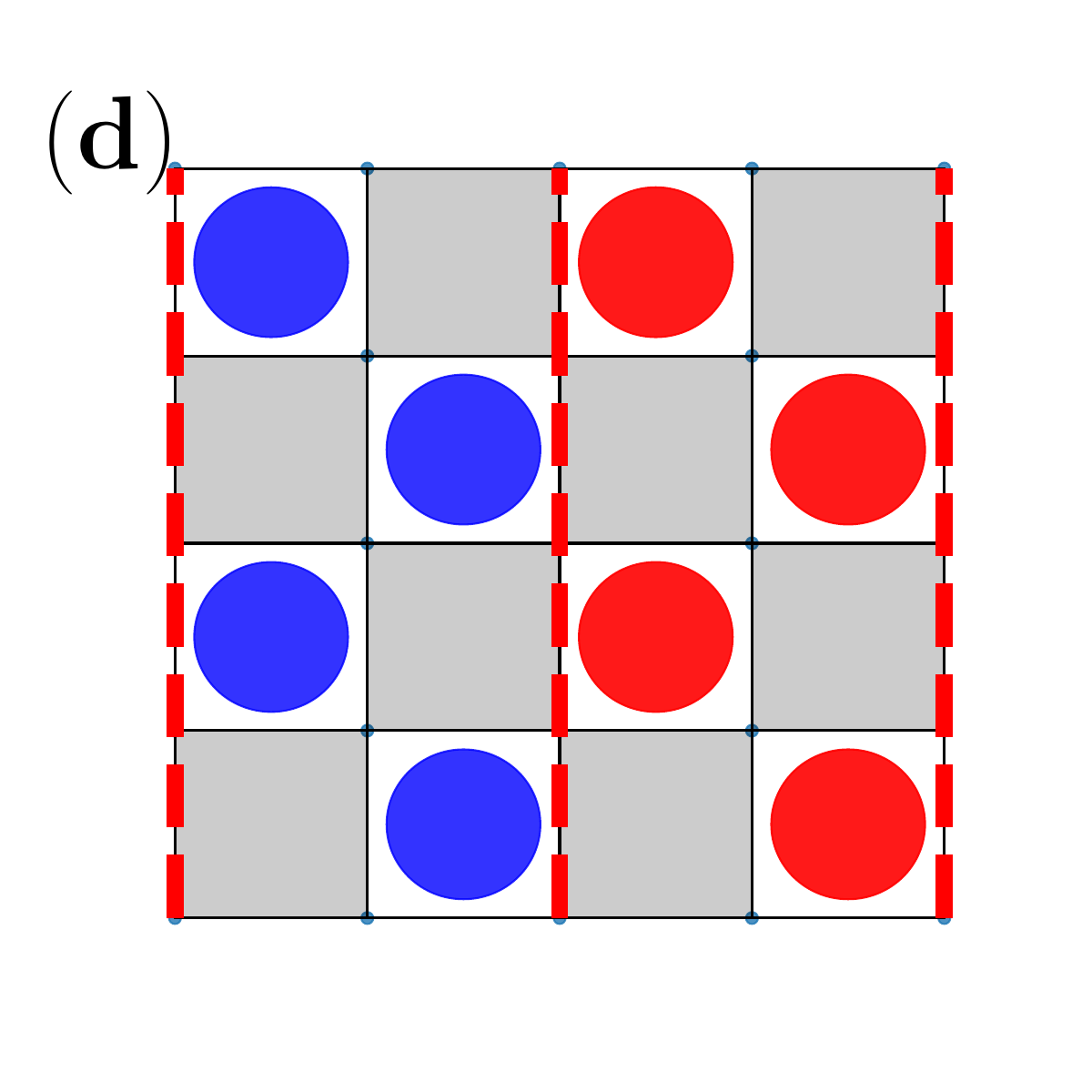}
    \includegraphics[width=0.18\textwidth]{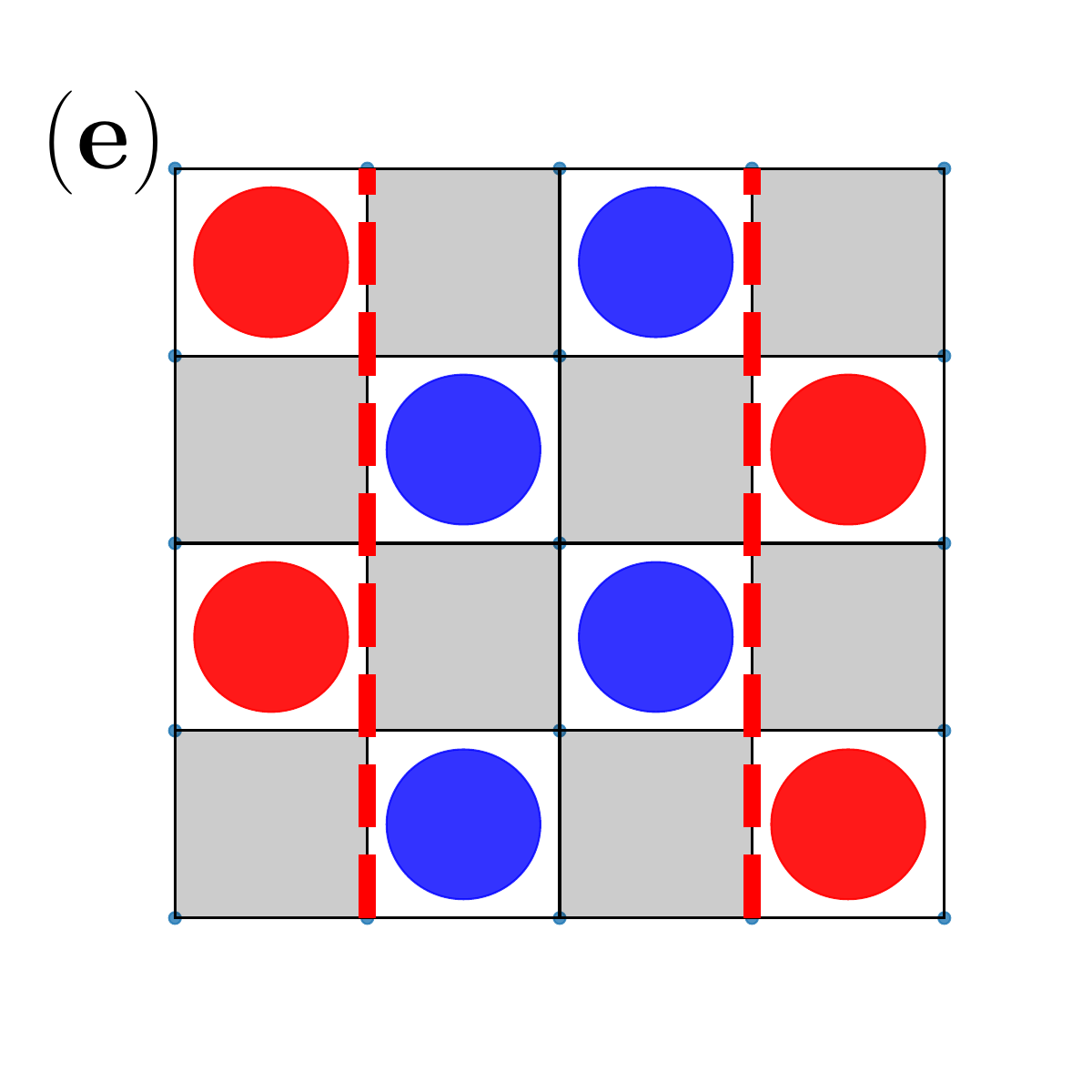}
    \includegraphics[width=0.22\textwidth]{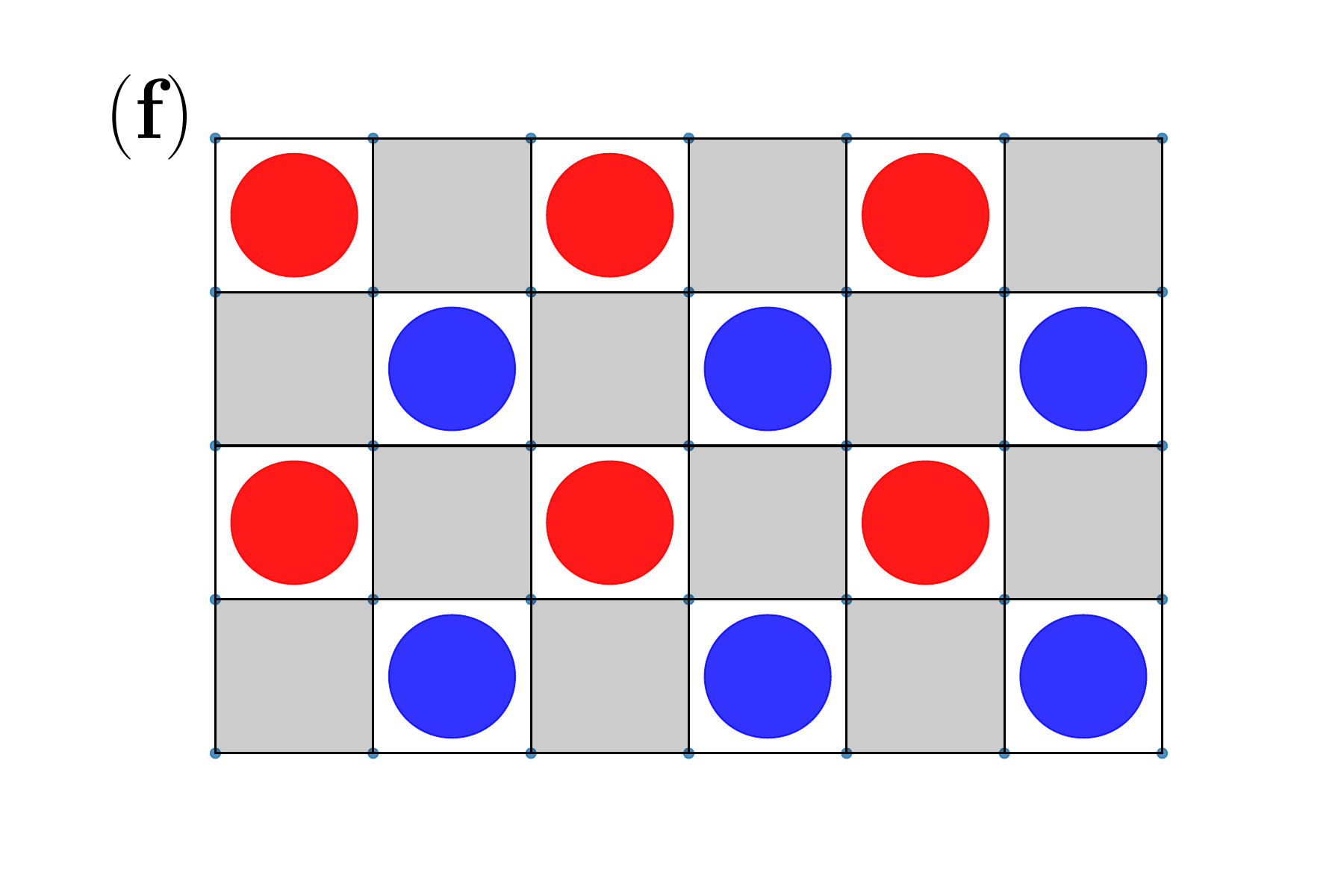}
    \includegraphics[width=0.22\textwidth]{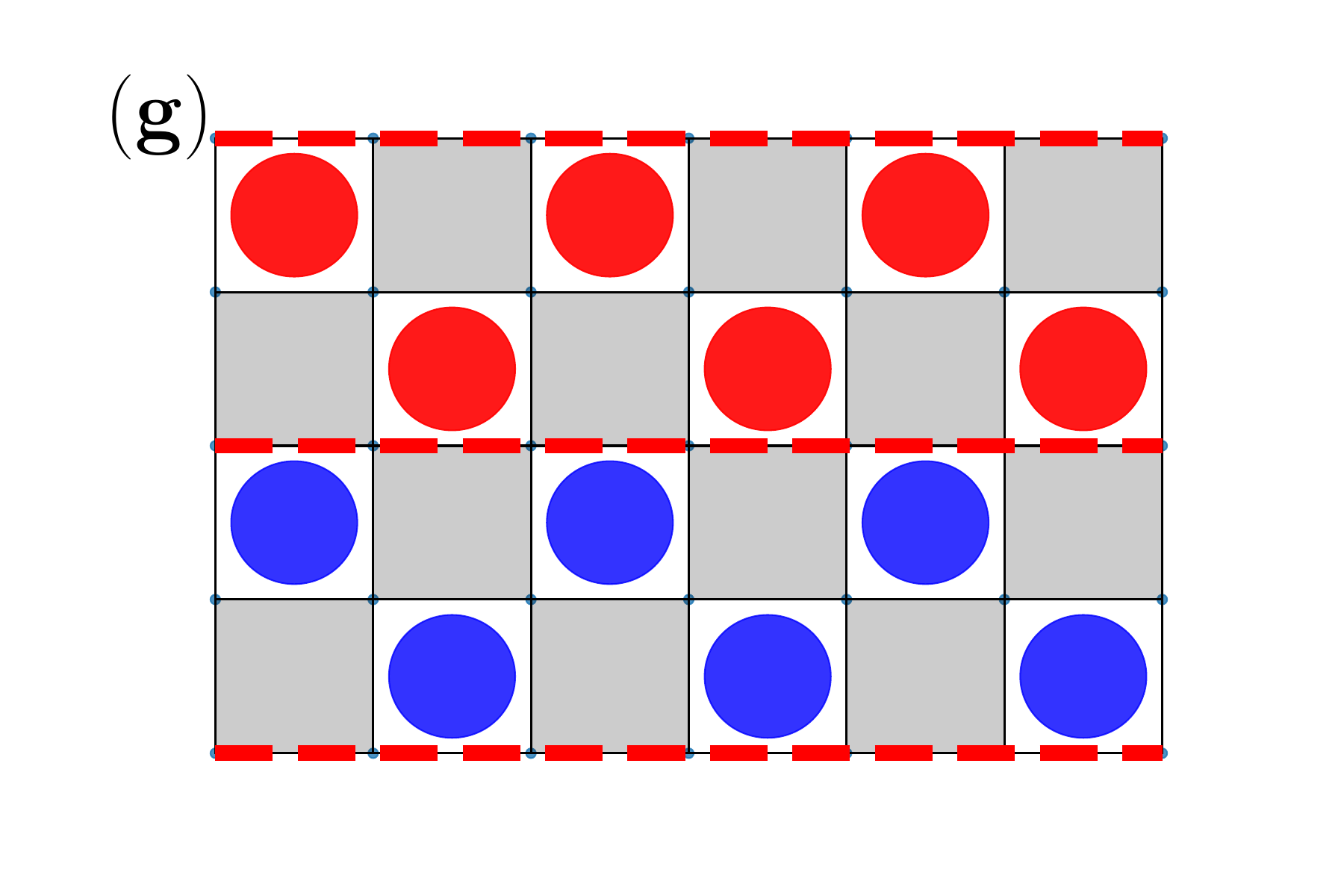}
    \includegraphics[width=0.22\textwidth]{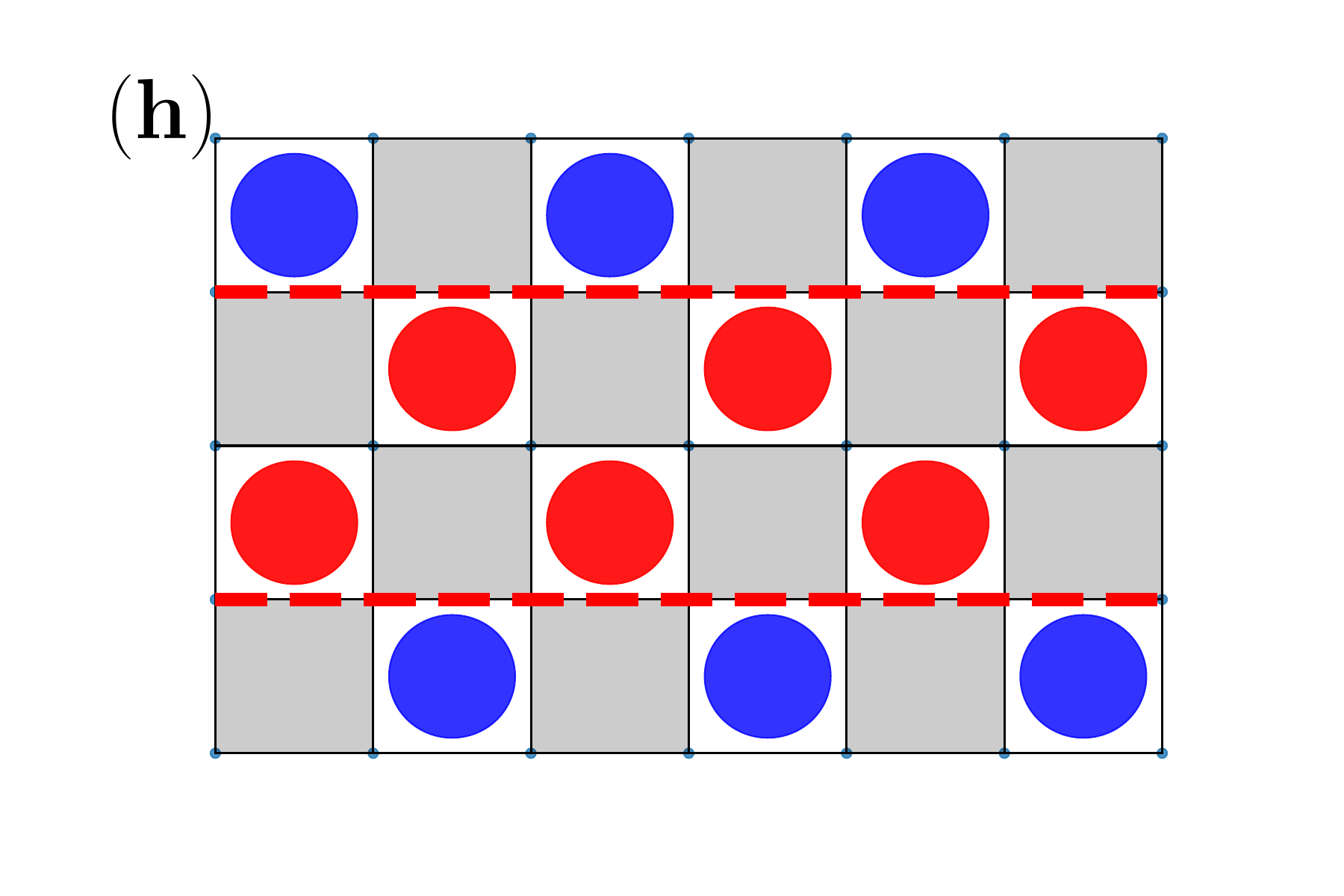}

    \caption{Base states from which the set $\{\ket{f_i}\}_n$ can be obtained shown here. 
    Panels (a) to (e) are for a $(4,4)$ lattice while panels (f) to (h) are for a $(6,4)$ 
    lattice. Clockwise (anti-clockwise) circulation of electric fluxes on a plaquette is 
    shown by a red (blue) circle as in Fig.~\ref{fig:qlmbasefig}. The white (gray) plaquettes 
    denote active (inactive) plaquettes with $\Opots=1 (0)$ as in Fig.~\ref{fig:shortsinglets}. 
    Horizontal (vertical) partitions of width $2$ are shown using dotted red lines as in 
    Fig.~\ref{fig:shortsinglets}. }
    \label{fig:base-states}
\end{figure}

 We now address the construction of the set $\{\ket{f_i}\}_n$. For this, we need to start with 
certain "base states" which are Fock states that satisfy the sublattice constraint and repeatedly 
act $\Okin$ on those states to get other Fock states consistent with the sublattice constraint. 
Most of the subsequent states are produced from the base state where plaquettes are arranged in an 
alternate clockwise and anticlockwise flippable pattern on the active sublattice as shown in 
Fig.~\ref{fig:base-states}($a$) for $(4,4)$ lattice and Fig.~\ref{fig:base-states}($f$) for 
$(6,4)$ lattice. Note that it is enough to consider any one of the two Fock states related by 
$\mathbb{C}$ for this base state. First we put the base state in $\{\ket{f_i}\}_n$. When $\Okin$ 
acts on this base state, new Fock states are generated out of which only some are consistent with 
the sublattice constraint, and we only insert these additional Fock states in $\{\ket{f_i}\}_n$. 
We then act $\Okin$ on these newly added Fock states to generate more Fock states and consider only 
the states consistent with the sublattice constraint out of these newly generated states to insert in 
$\{\ket{f_i}\}_n$. At this stage, we avoid adding the Fock states which are already present in 
$\{\ket{f_i}\}_n$. We repeat this recursive process until there is no new Fock state to be added to 
the set $\{\ket{f_i}\}_n$. This procedure leaves out a few Fock states from the set $\{\ket{f_i}\}_n$. 
These "base states", which are consistent with the sublattice constraint, have the property that the 
action of $\Okin$ on them produces no Fock state in the set $\{\ket{f_i}\}_n$ and, hence, these need 
to be inserted separately in $\{\ket{f_i}\}_n$. There is a simple rule to generate these extra base 
states to complete the set $\{\ket{f_i}\}_n$. We first divide the lattice into a close packing of 
parallel non-overlapping horizontal or vertical partitions, each of width $2$ and then arrange all 
the active plaquettes contained in a partition in a clockwise or anticlockwise manner. Once active
plaquettes in a partition are arranged in a clockwise (anticlockwise) manner, its neighboring partitions 
must have active plaquettes with the opposite circulation. Lastly, the number of partitions with 
clockwise active plaquettes must equal that of partitions with anticlockwise active plaquettes in 
such a base state.  Fig.~\ref{fig:base-states}($b$) to Fig.~\ref{fig:base-states}($e$) show such base 
states for a $(4,4)$ lattice and Fig.~\ref{fig:base-states}($g$) and Fig.~\ref{fig:base-states}($h$) 
show such base states for the $(6,4)$ lattice. Note that only half of these base states are shown 
here with the other half easily generated by applying the charge conjugation operator $\mathbb{C}$.

\section{Parent Hamiltonian for sublattice scars}
\label{sec:parent}

 The sublattice scars discussed in the previous sections are mid-spectrum eigenstates of 
$\mathcal{H}_{\mathrm{RK}}$ for any $\lambda \sim O(1)$. We can ask whether a parent Hamiltonian can 
be written for which these states are, in fact, ground states. 
One motivation for this is to substantiate the expectation, that although these scar 
states occur as high-energy excited states of a lattice Hamiltonian, they should not be 
dismissed as cut-off effects. On the contrary, as we show next, such states can also be realized
in the low-energy physics of certain (gauge invariant) Hamiltonian, and therefore possibly survive 
the continuum limit. We do not however, have a rigorous proof of the latter and leave it for 
future investigation.

The following {\em long-ranged} Hamiltonian:
\begin{eqnarray}
  \mathcal{H}_{\mathrm{LR}} &=& \frac{1}{N_p}(\Okin)^2 +  c \sum_\square (-1)^\square \Opots \nonumber \\
  &=& \frac{1}{N_p} \sum_{\square_i, \square_j} (\Okins)_i (\Okins)_j \nonumber \\
  &+&  c \sum_\square (-1)^\square \Opots 
  \label{eq:longrangedH}
\end{eqnarray}
which consists of an {\em all-to-all} two-plaquette interaction of the form $(\Okins)_i (\Okins)_j$,
where the indices $i$ and $j$ run over all plaquettes, and another short-ranged staggered term involving 
$\Opots$. The normalization of $1/N_p$ for the first term ensures that both the terms scale extensively 
with system size. Lastly, $c \neq 0$ is an arbitrary real parameter (which we choose to be O(1)).  
\begin{figure}
    \centering
    \includegraphics[scale=0.25]{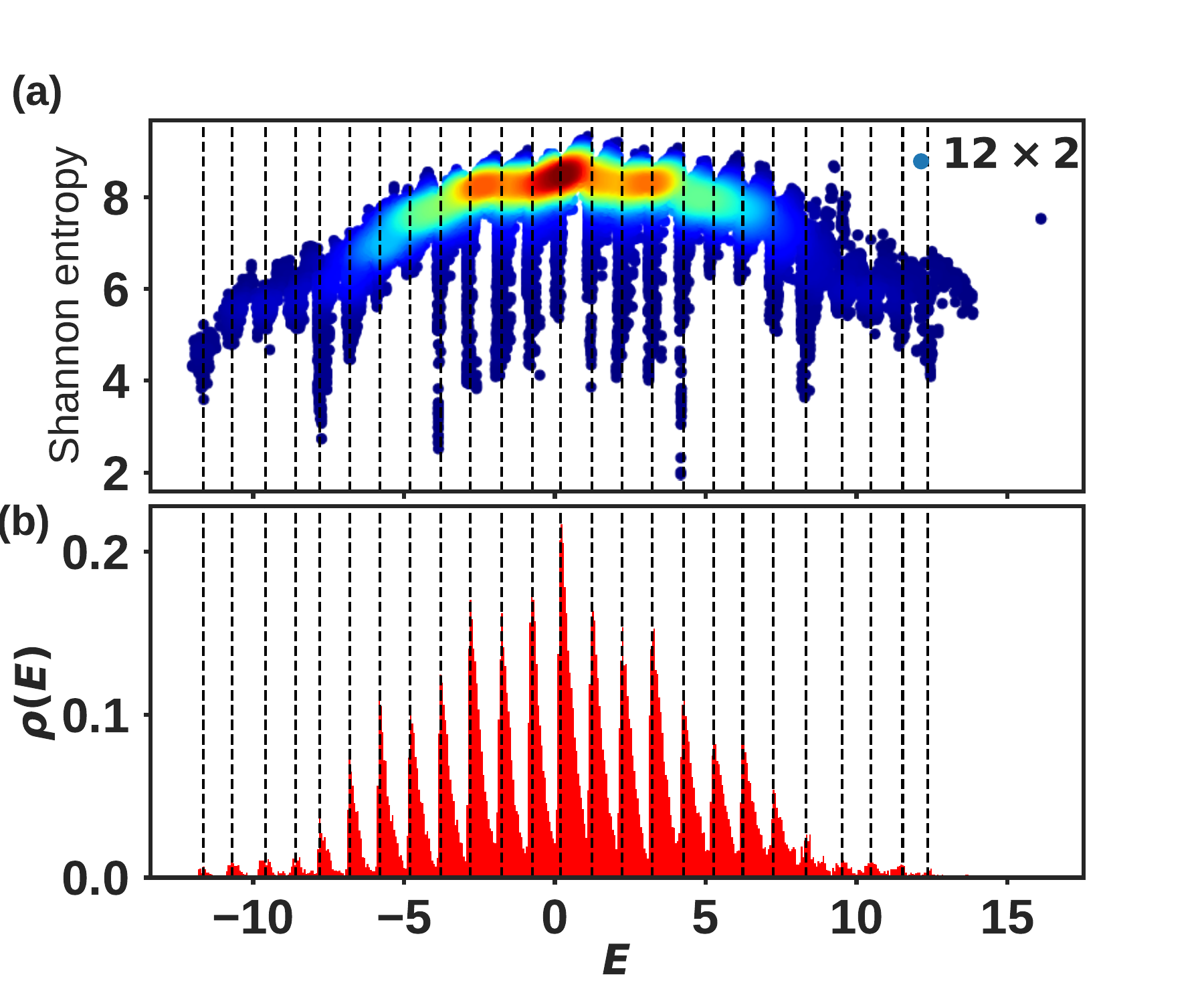}
    \caption{(a) Behavior of the Shannon entropy shown for all the eigenstates of $\mathcal{H}_{\mathrm{LR}}$ 
    for a system of size $(12,2)$. (b) Density of states, $\rho(E)$, plotted as a function of energy for the 
    same system. Both calculations use $c=1.0$. In (b), $512$ bins have been taken to find density of states. 
    Vertical dotted lines are plotted at the positions of local maxima of DOS by visual inspection. In (a), 
    the density of states is indicated by a color map where warmer color corresponds to higher density of states.}
    \label{fig:DOS_LRPH}
\end{figure}

For $c>0$ ($c<0$), all sublattice scars with $\Okin=0$ and active plaquettes on one (the other) sublattice 
become exact ground states of $\mathcal{H}_{\mathrm{LR}}$ since both $(\Okin)^2$ and 
$c\sum_\square (-1)^\square \Opots$ are minimized simultaneously. In fact, it is enough to focus on one 
particular sign of $c$. While sublattice scars with active plaquettes on one sublattice become ground 
states with energy $E_0 = -|c|\frac{N_p}{2}$, the other set of sublattice scars (where the active plaquettes 
reside on the other sublattice) become exact eigenstates of $\mathcal{H}_{\mathrm{LR}}$ with energy 
$-E_0 = |c| \frac{N_p}{2}$. The sublattice scars with $\Okin = \pm 2$ become \emph{degenerate} eigenstates 
of $\mathcal{H}_{\mathrm{LR}}$ with energies $E_{s,\mp} = (\frac{4}{N_p}) \mp |c|\frac{N_p}{2}$, depending 
on which sublattice the active plaquettes belong to.

 The triangle relation (Eq.~\ref{eq:trianglerelation}) connecting sublattice scars with $\Okin=\pm 2$ with 
particular sublattices scars with $\Okin=0$ has a nice interpretation in terms of the states of 
$\mathcal{H}_{\mathrm{LR}}$. The sublattice scar $\ket{\psi_{s,0}}$, composed of Fock states with unequal 
number of clockwise and anticlockwise flippable active plaquettes, is a ground state of
$\mathcal{H}_{\mathrm{LR}}$ (with an appropriate sign of $c$ to choose the sublattice). The state 
$\ket{\psi_{s,+2}} - \mathcal{C}_\alpha \ket{\psi_{s,+2}}$ is an eigenstate of $\mathcal{H}_{\mathrm{LR}}$ 
with energy $E_{s,-} =  (\frac{4}{N_p}) - |c| \frac{N_p}{2}$. Note that $\ket{\psi_{s,+2}}$ and 
$\mathcal{C}_\alpha \ket{\psi_{s,+2}}$ have different energies with respect to $\mathcal{H}_{\mathrm{RK}}$, 
so their difference is not an eigenstate of $\mathcal{H}_{\mathrm{RK}}$. Since the operator $\mathcal{O}$ 
leads to a sign change (no sign change) for Fock states with 
$(\frac{N_p}{4} + 1, \frac{N_p}{4} - 1)$ ($(\frac{N_p}{4} - 1, \frac{N_p}{4} + 1)$) clockwise and anti-clockwise 
active plaquettes, it is clear that $\mathcal{O}^2=1$. Using this, we can rewrite Eq.~\ref{eq:trianglerelation} 
as 
\begin{equation}
  \left( \ket{\psi_{s,+2}} - \mathcal{C}_\alpha \ket{\psi_{s,+2}} \right) \propto \mathcal{O} \ket{\psi_{s,0}}
  \label{eq:trianglerelation2}
\end{equation}
where $\mathcal{O}$ can now be interpreted to creates a finite-energy excitation, with an energy of 
$\Delta E = \frac{4}{N_p}$, when acting on a ground state of $\mathcal{H}_{\mathrm{LR}}$. However, 
given the form of $\mathcal{O}$, it does not seem possible to write it as a sum of local (in space) operators.

We have verified numerically using ED on finite $(L_x,L_y)$ lattice that the ground states (with energy $-E_0$), 
excited states with energy $+E_0$, and energies $E_{s,\mp}$ are all sublattice scars. ED also reveals a rich 
structure in the spectrum of $\mathcal{H}_{\mathrm{LR}}$. We display the data for a system with dimension 
$(12,2)$ with $c=1$ in Fig.~\ref{fig:DOS_LRPH}. Fig.~\ref{fig:DOS_LRPH}(a) shows Shannon entropy for each 
eigenstate $\ket{\Psi}$ which can be calculated using $-\sum_\alpha |\Psi_\alpha|^2 \ln |\Psi_\alpha|^2$ , 
where $\ket{\Psi} =\sum_\alpha \Psi_\alpha \ket{\alpha}$ when the eigenstate is expressed in the computational 
basis $\ket{\alpha}$. The Shannon entropy shows several prominent dips as a function of energy $E$ indicating 
the presence of several eigenstates which are much more localized in the Hilbert space compared to neighboring 
eigenstates. Interestingly, the density of states, $\rho(E)$, extracted from the eigenvalues of 
$\mathcal{H}_{\mathrm{LR}}$ display a rather intricate structure as well with several local maxima 
(Fig.~\ref{fig:DOS_LRPH}(b)). The positions of these local maxima in $\rho(E)$ seem to be strongly correlated 
to the appearance of anomalous dips in the Shannon entropy.

Finally, from ED on finite lattices, the zero energy states of $\mathcal{H}_{\mathrm{LR}}$ {\em stay unchanged} 
(apart from mixing with each other) as a function of $c$. This implies that these must be simultaneous eigenstates
of $\Okin$ and  $c \sum_\square (-1)^\square \Opots$. We will discuss these anomalous zero modes, which are 
distinct from the sublattice scars, in the next section.

\section{More quantum scars from zero modes}
\label{sec:moreScars}
\begin{figure}
     \centering
     \includegraphics[width=\linewidth]{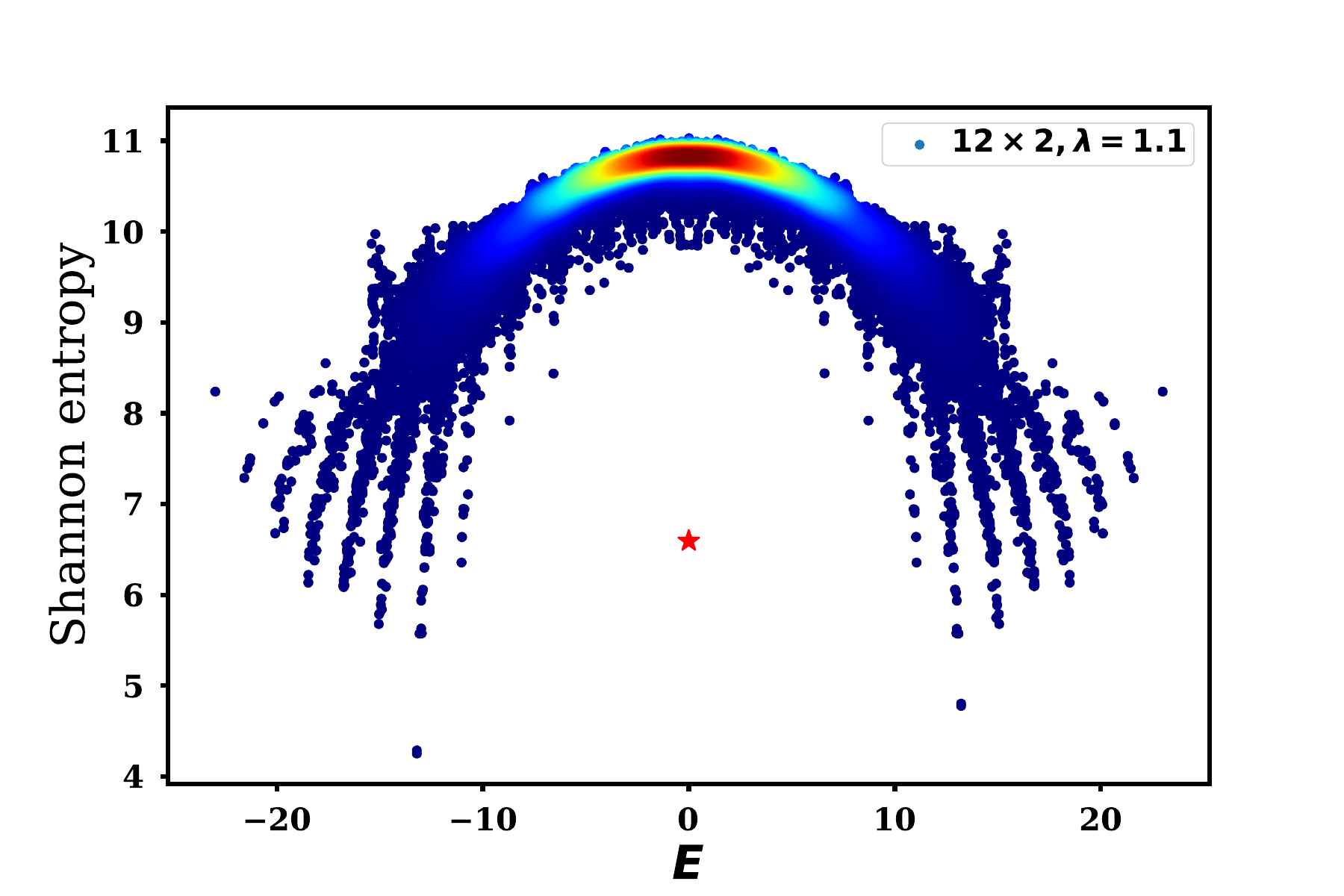}
      \includegraphics[width=\linewidth]{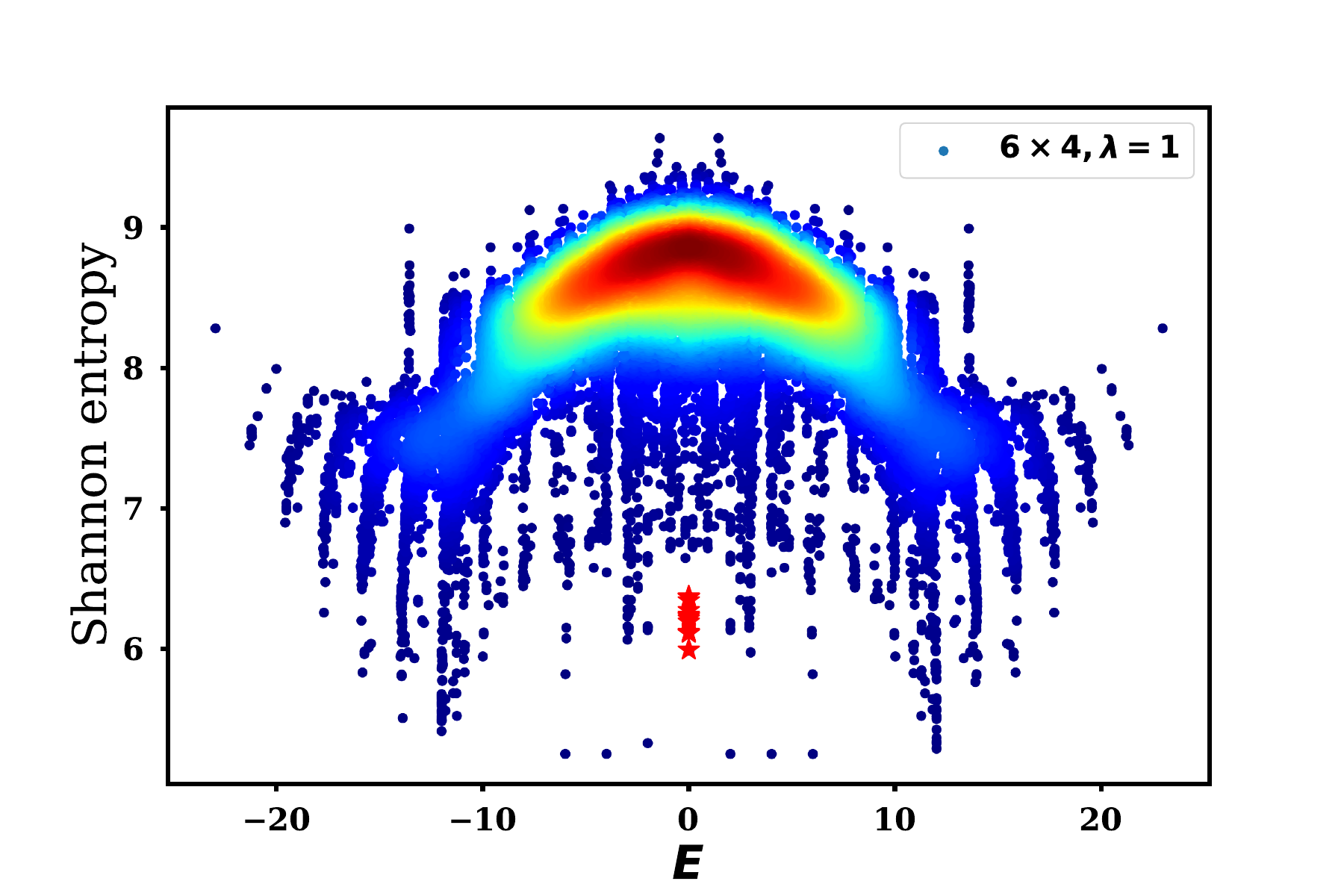}
      \caption{Shannon entropy for all the eigenstates of $\mathcal{H}_{\mathrm{st}}$ (Eq.~\ref{eq:H2}) 
      plotted as a function of energy $E$ with panel (a) showing data for system of size $(12,2)$ and 
      panel (b) showing data for $(6,4)$. In both plots, the simultaneous zero modes of $\Okin$ and 
      $\sum_{\square} (-1)^\square \Opots$ are shown using a different font and color (red stars). The 
      density of states is indicated by the same color map in both panels where warmer color corresponds 
      to higher density of states.}
      \label{fig:Hstscars}
  \end{figure}
 We now discuss a different variety of quantum many-body scars that are again composed of the null space 
 of $\Okin$ but have different properties from the sublattice scars. While the sublattice scars are 
 simultaneous eigenstates of the non-commuting operators $\Okin$ and $\sum_\square \Opots$, these scars 
 are instead simultaneous \emph{zero modes} of $\Okin$ and $\sum_\square (-1)^\square\Opots$. These 
 anomalous states are thus exact zero modes of $\mathcal{H}_{\mathrm{st}}$ that stay unchanged as a 
 function of $\lambda$. We refer the reader to Ref.~\onlinecite{Samudra2023} for a similar scarring 
 mechanism in a spin chain. As discussed in a previous section, $\mathcal{H}_{\mathrm{st}}$ satisfies 
 an index theorem at any value of $\lambda$ and, hence, its spectrum has an $E$ to $-E$ symmetry as well 
 as exponentially many (in system size) zero modes (see Table~\ref{tab:zromodes} and Fig.~\ref{fig:Nzeroindex}). 
 However, since $\Okin$ and $\sum_\square (-1)^\square\Opots$ do not commute with each other, these zero modes, 
 apart from the ones that are simultaneous eigenstates of both the terms, keep changing in a non-trivial 
 fashion. In fact, any typical zero mode of $\mathcal{H}_{\mathrm{st}}$ is expected to satisfy ETH and 
 locally mimic a featureless infinite temperature state. However, the zero modes that stay unchanged with 
 $\lambda$ are expected to violate the ETH~\cite{Banerjee2021,Samudra2023}.
     
 The anomalous nature of these simultaneous zero modes can be clearly demonstrated by calculating the 
Shannon entropy of all the eigenstates of $\mathcal{H}_{\mathrm{st}}$ (Eq.~\ref{eq:H2}) for $\lambda \sim O(1)$ 
using ED on finite lattices. Fig.~\ref{fig:Hstscars} displays the data for $(12,2)$ at $\lambda=1.1$ and for 
$(6,4)$ at $\lambda=1.0$. It is clear from both panels that the simultaneous zero modes have anomalously low 
Shannon entropy compared to the typical values for neighboring eigenstates, demonstrating that these states 
do violate the ETH. From both plots, it is also clear that there may be other anomalous high-energy states 
in the spectrum of $\mathcal{H}_{\mathrm{st}}$ but we only focus on the ETH-violating zero modes here. 
     
   \begin{table}[h]
   	\begin{tabular}{|c|c|}
   		\hline
   		Lattice & Anomalous zero modes of $\mathcal{H}_{\mathrm{st}}$\\
   		\hline
   		$L_x\times2$ & 2 \\
   		\hline
   		$4\times4$ & 8 \\
   		\hline
   		$6\times4$ & 14 \\
   		\hline
   	\end{tabular}
    \caption{Scaling of the number of anomalous zero modes of $\mathcal{H}_{\mathrm{st}}$ as a function of 
    system size extracted from ED.} 
    \label{table:zeroModesLRPH}
   \end{table}

   The number of these simultaneous zero modes is given in Table.~\ref{table:zeroModesLRPH}. Just like in 
the case of sublattice scars, their number stays fixed with $L_x$ in the thin torus limit of $L_y=2$, while 
it increases with $L_x$ for wider ladders. These quantum scars are eigenstates of the charge conjugation, 
$\mathbb{C}$, with half of them possessing $\mathbb{C}=+1$ ($\mathbb{C}=-1$). For $(L_x,2)$ systems, the 
unique scar with $\mathbb{C}=1$ ($\mathbb{C}=-1$) is also a momentum eigenstate with $(k_x,k_y)=(0,0)$ 
($=(\pi,\pi)$) with respect to translations by one lattice unit in both directions.   

\begin{figure}[h]
        \centering
        \includegraphics[scale=0.12]{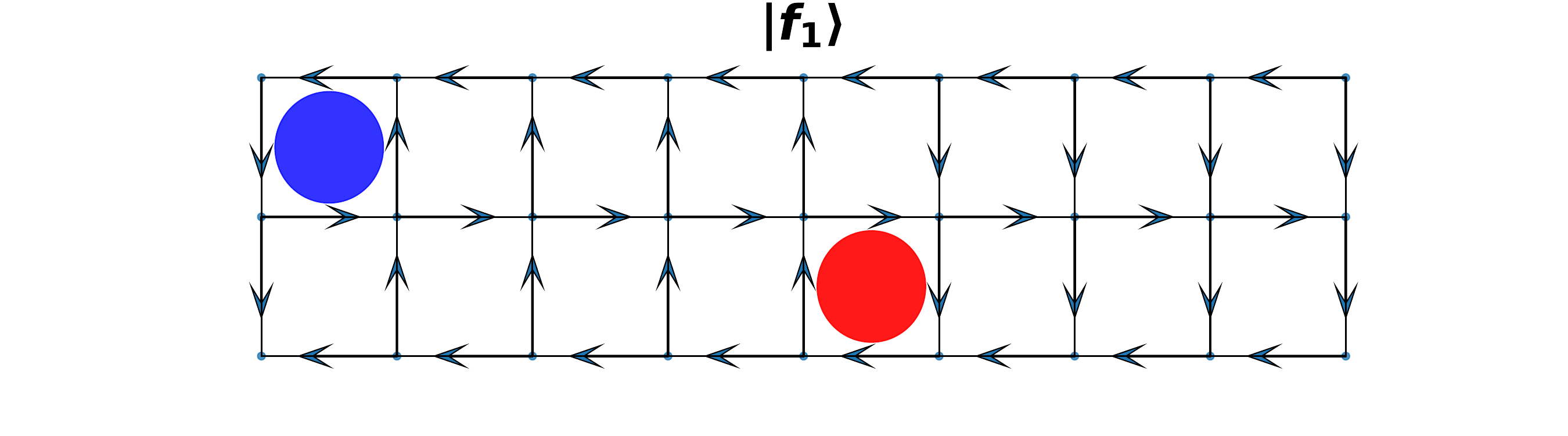}
        \includegraphics[scale=0.12]{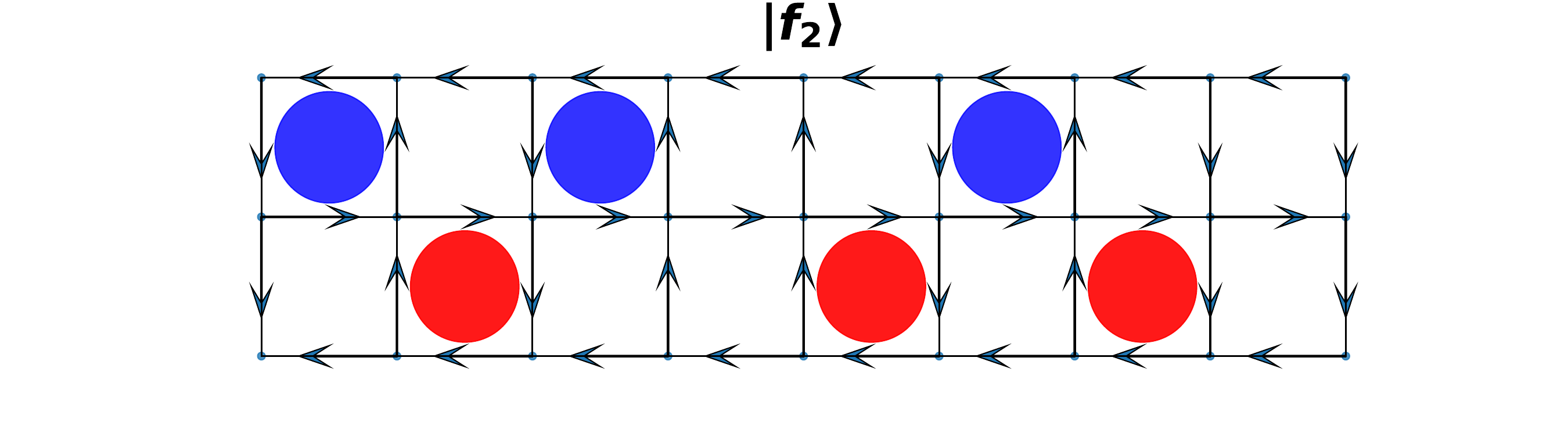}
        \includegraphics[scale=0.12]{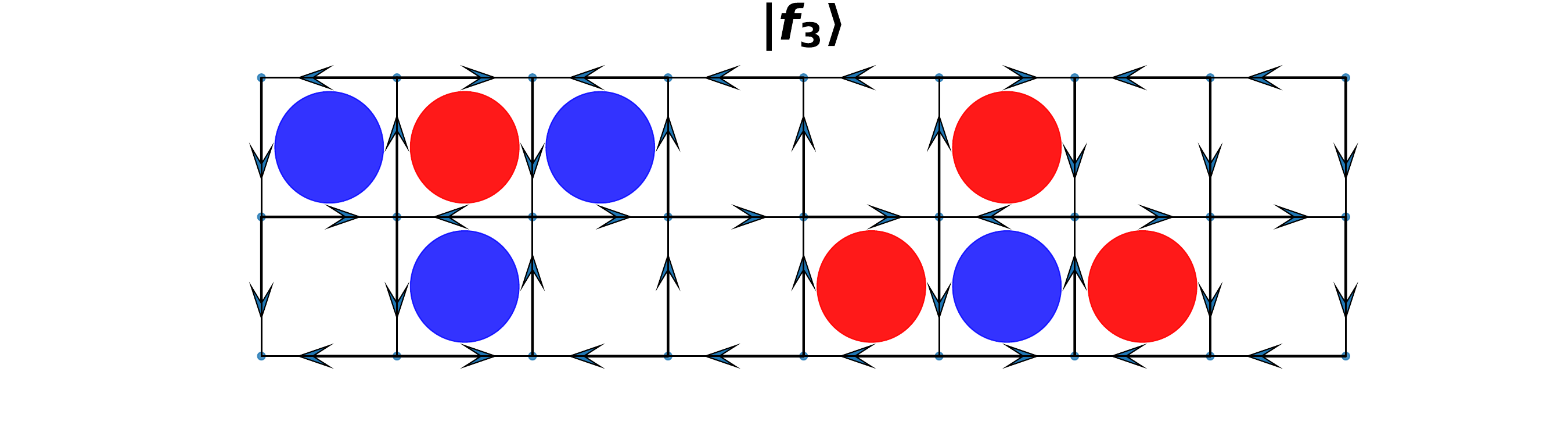}
        \includegraphics[scale=0.12]{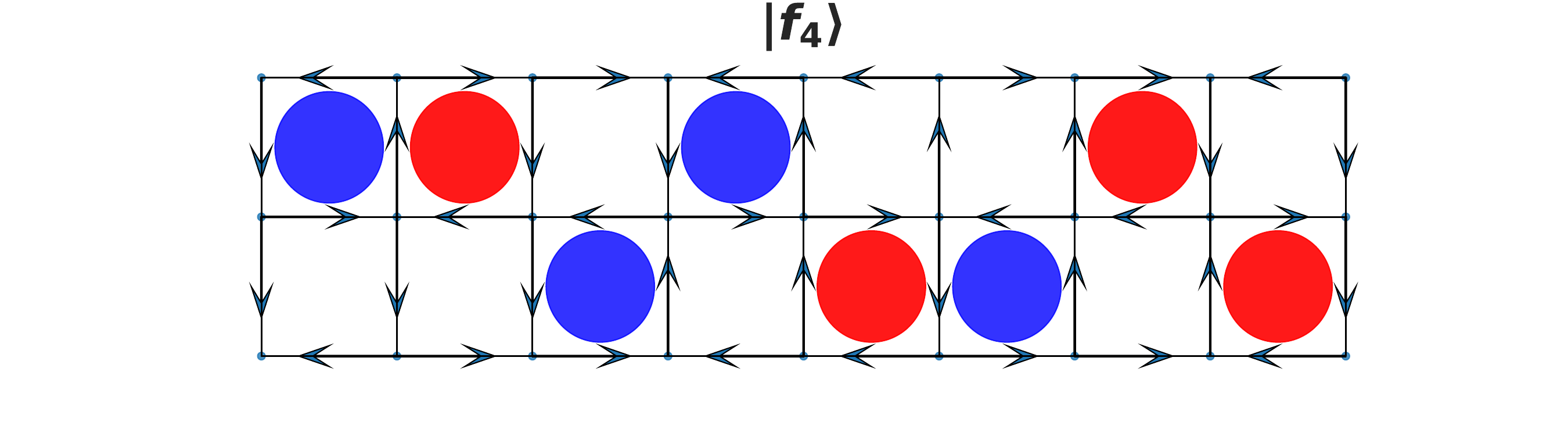}
        \includegraphics[scale=0.12]{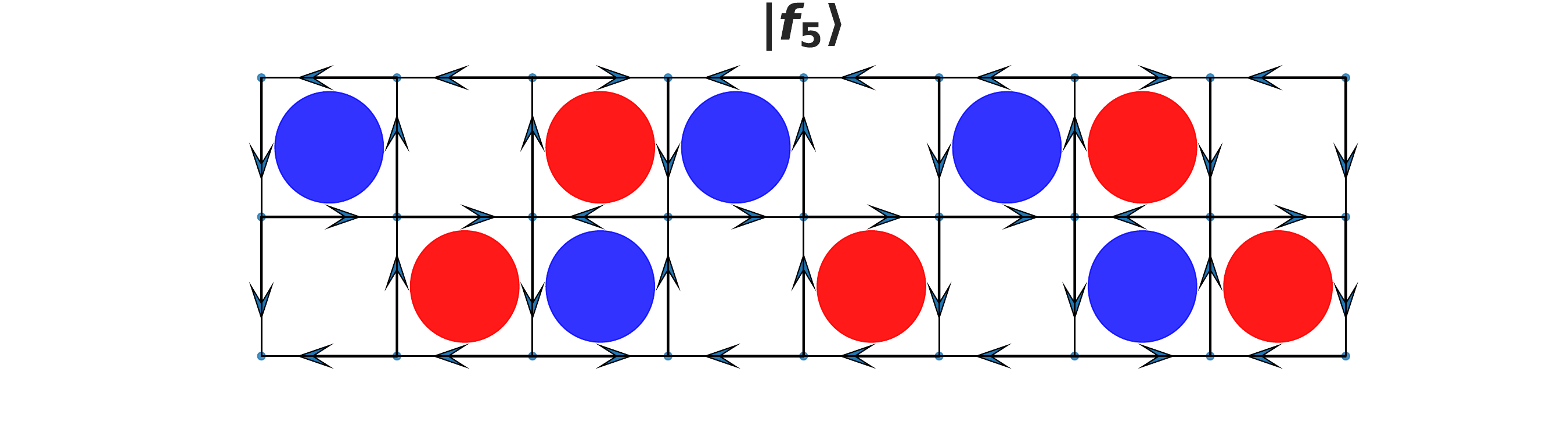}
        \includegraphics[scale=0.12]{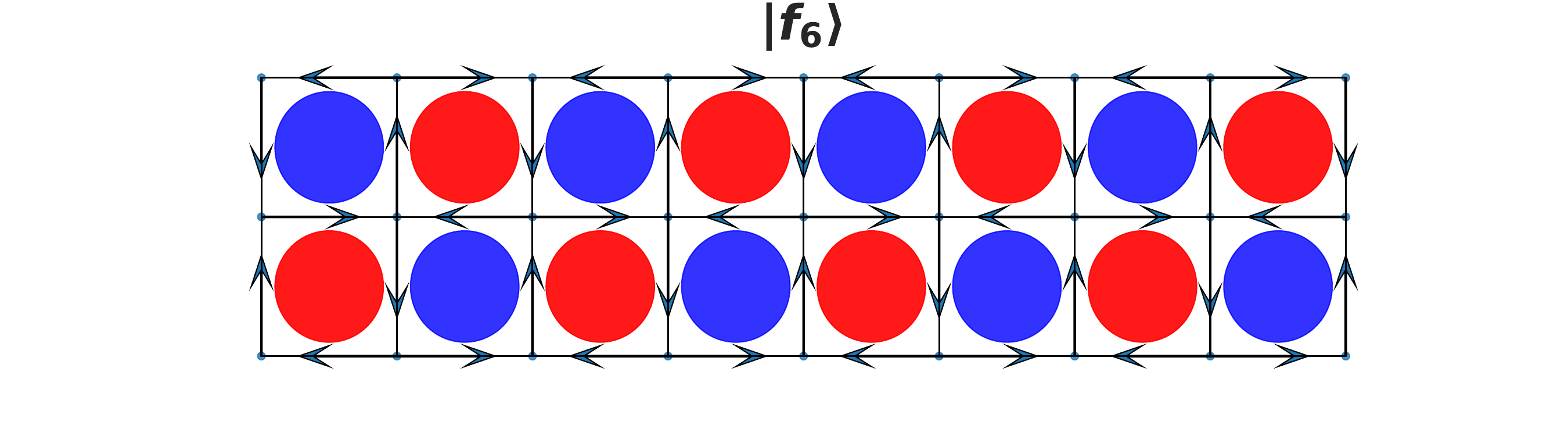}
        \caption{The representative states ($\ket{f_i}$) that build the two anomalous zero modes of 
        $\mathcal{H}_{\mathrm{st}}$ for a system of dimension $(8,2)$. The electric fluxes are shown 
        for each link of each of the representative states by arrows. Clockwise (anticlockwise) circulation 
        of electric fluxes on a plaquette is shown by a red (blue) circle as in Fig.~\ref{fig:qlmbasefig}.}
        \label{fig:scar8times2Hst}
    \end{figure}

While we do not yet have an analytic understanding of these quantum scars, we show the form of these states 
in the thin torus limit of $L_y=2$ for $L_x=8$. The representative Fock states (denoted as $\ket{f_i}$ where 
$i$ ranges from $1$ to $6$) that contribute to create the two simultaneous zero modes for a $(8,2)$ lattice 
are shown in Fig.~\ref{fig:scar8times2Hst}.  Given any of the representative states, $\ket{f_i}$, one can build 
a basis state 
$\ket{f_i(k_x,k_y)}=\frac{1}{\sqrt{N_i}}\sum_{x=1}^{L_x}\sum_{y=1}^{L_y}e^{-i(k_x x + k_y y)}T_Y^y T_X^x \ket{f_i}$ 
that carries a well-defined momentum $(k_x,k_y)$ with respect to translations by one lattice unit in $(x,y)$ 
implemented by the operators $(T_X,T_Y)$. Note that the renormalization $N_i$ represents the number of distinct 
Fock states that be obtained from $\ket{f_i}$ by translations and equals $N_i=6$ for $i=1$ to $5$ and $N_6=2$. 
The two anomalous zero modes of $\mathcal{H}_{\mathrm{st}}$ for $(8,2)$ can now be compactly written as follows:
 \begin{eqnarray}
   \ket{\Psi_{s,+1}} &=&\frac{1}{\sqrt{41}}[\sqrt{3} \ket{f_1(0,0)}]+\sqrt{3}\ket{f_2(0,0)}
   - \sqrt{3} \ket{f_3(0,0)}  \nonumber \\
     &-& \sqrt{3}\ket{f_4(0,0)} +\sqrt{3} \ket{f_5(0,0)} + \ket{f_6(0,0)}] \nonumber \\
    \ket{\Psi_{s,-1}} &=&\frac{1}{\sqrt{41}}[\sqrt{3} \ket{f_1(\pi,\pi)}] - \sqrt{3} \ket{f_2(\pi,\pi)}
    + \sqrt{3} \ket{f_3(\pi,\pi)}  \nonumber \\
  &-& \sqrt{3} \ket{f_4(\pi,\pi)} -\sqrt{3} \ket{f_5(\pi,\pi)} - \ket{f_6(\pi,\pi)}]
         \label{eq:scars8times2}
     \end{eqnarray}
   where $\ket{\Psi_{s,+1}}$ carries momentum $(k_x,k_y)=(0,0)$ and $\mathbb{C}=+1$ while 
   $\ket{\Psi_{s,-1}}$ has $(k_x,k_y)=(\pi,\pi)$ and $\mathbb{C}=-1$. From the structure of 
   the representative Fock states in Fig.~\ref{fig:scar8times2Hst}, it is also clear that 
   these states \emph{do not} have a sublattice structure of active and inactive plaquettes 
   in real space.
   
\section{Conclusion and outlook}
\label{sec:conclusions}
  In conclusion, we have considered two $U(1)$ quantum link gauge theory Hamiltonians in their 
spin-$\frac{1}{2}$ representation on $(L_x,L_y)$ square lattices with periodic boundary conditions 
in both directions and even $L_x,L_y$. We specifically target the largest superselection sector with 
zero charge at each site and zero winding numbers in both directions. Both the link models are composed 
of plaquette operators, $\Okins$ and $\Opots$, defined on elementary plaquettes of the lattice where 
$\Okins$ changes a clockwise circulation of electric fluxes on a plaquette to anti-clockwise and 
vice-versa while $\Opots$ acts as a counting operator for such a plaquette. While the first Hamiltonian 
that we consider, $\mathcal{H}_{\mathrm{RK}}=-\sum_\square \Okins+\lambda \sum_\square\Opots$ 
(Eq.~\ref{eq:H1}), is a standard Rokhsar-Kivelson model, the second Hamiltonian, 
$\mathcal{H}_{\mathrm{st}} = -\sum_\square \Okins+\lambda \sum_\square (-1)^\square \Opots$ 
(Eq.~\ref{eq:H2}) has a staggered potential term that depends on the sublattice location of the 
plaquette. Both models represent non-integrable Abelian lattice gauge theories without dynamical 
matter. Both theories become identical when $\lambda=0$ where the Hamiltonian, which can be represented 
as $\Okin=-\sum_\square \Okins$, only consists of off-diagonal terms in the electric flux basis. 
In this limit, the system supports an exponentially large number (in system size) of exact mid-spectrum 
zero modes that are protected by an index theorem. The index theorem is immediately lifted when 
$\lambda \neq 0$ for $\mathcal{H}_{\mathrm{RK}}$ while it stays preserved for any $\lambda$ for 
$\mathcal{H}_{\mathrm{st}}$. The zero modes of $\Okin$ as well as $\mathcal{H}_{\mathrm{st}}$ are 
expected to locally mimic a featureless infinite temperature thermal state from the eigenstate 
thermalization hypothesis.    

  We show the existence of several anomalous high-energy eigenstates, that violate the eigenstate 
thermalization hypothesis, in both these quantum link gauge theory Hamiltonians. We dub one class of 
these anomalous states as sublattice scars. These sublattice scars, $\ket{\psi_s}$, are highly structured 
in terms of $\Opots$ with $\Opots \ket{\psi_s} = \ket{\psi_s}$ for all elementary plaquettes on one 
sublattice and $\Opots \ket{\psi_s}=0$ on the elementary plaquettes that belong to the other sublattice. 
Furthermore, these states are eigenstates of $\Okin$, with eigenvalues $0$ or $\pm 2$. A class of 
sublattice scars with $\Okin=0$ have a simple representation in terms of coverings of emergent dimers 
(singlets) and their number scales as $O(2^{L_x/2+1}+2^{L_y/2+1}-4)$ for $L_x,L_y \gg 1$, thus showing 
their presence even in two dimensions. These short singlet scars explain all sublattice scars that occur 
in the thin-torus limit of $L_y=2$ and $L_x$ arbitrary. However, wider systems with $L_y \geq 4$ have 
several sublattice scars that are beyond this description in terms of singlets. We also find sublattice 
scars with $\Okin=\pm 2$ for $L_y \geq 4$ and demonstrate a non-trivial ``triangle relation'' between 
sublattice scars with $\Okin=+2$, $\Okin=-2$ and $\Okin=0$. The analysis of the structure of the 
non-singlet scars was greatly aided by a numerical approach which directly focused on the relevant state 
space for these high-energy states, instead of usual exact diagonalization techniques which require 
the full Hilbert space to construct high-energy eigenstates. We further discuss a long-ranged parent 
Hamiltonian, $\mathcal{H}_{\mathrm{LR}}$ (Eq.~\ref{eq:longrangedH}), which gives all sublattice scars 
with $\Okin=0$ as unique ground states and sublattice scars with $\Okin=\pm 2$ as finite energy 
excitations. The triangle relation can be interpreted as a \emph{quasiparticle} operator acting on 
certain ground states of this long-ranged Hamiltonian to give finite energy excitations. Apart from 
these sublattice scars, we find additional anomalous zero modes of $\Okin$ that are also exact zero 
modes of $\sum_{\square} (-1)^\square \Opots$, and hence anomalous zero modes of $\mathcal{H}_{\mathrm{st}}$ that 
do not change with coupling $\lambda$. 

The results presented here immediately suggest several future avenues of study. One of the
conceptually challenging questions is to establish the presence of such anomalous eigenstates in the
limit when the lattice discretization is removed, i.e., directly in the continuum field theory. 
One way we have tried to motivate that this indeed
might be the case is to construct a different Hamiltonian, where such states occur in the ground state
manifold, indicating that the scarring phenomena resulting in the eigenstates reported here can exist
beyond a single Hamiltonian and energy window. However, a detailed analysis is required. It might be 
interesting to attempt the description of these quantum scars using the langugage of path integrals, 
using which it might be easier to establish their behaviour in the continuum limit.

It might also be worthwhile to explore if addition of local kinetic energy terms can also 
give rise to the singlet scar states as ground states. Moreover, the spectrum of $\mathcal{H}_{\mathrm{st}}$ 
seems to suggest the presence of several other anomalous high-energy states apart from the ones discussed here. 
We leave this as an interesting question for future studies. A deeper understanding of both the anomalous 
zero modes of $\mathcal{H}_{\mathrm{st}}$ and the non-singlet sublattice scars would be highly desirable. 
Additionally, the ground state physics, and possible phase transitions of both the 
the $\mathcal{H}_{\mathrm{LR}}$ and the $\mathcal{H}_{\mathrm{st}}$ would be an exciting line of future
research, given the recent interest of realizing lattice gauge theories on quantum simulator platforms.
Finally, another interesting question is to search for similar scars in spin-$S$, with $S \geq 1$, quantum link 
gauge theories without dynamical matter fields as well as in non-Abelian versions. 

\begin{acknowledgments}
We would like to acknowledge the computational resources provided by IACS and SINP. 
PS acknowledges support from: ERC AdG NOQIA; MCIN/AEI (PGC2018-0910.13039/501100011033,  CEX2019-000910-S/10.13039/501100011033, Plan National FIDEUA PID2019-106901GB-I00, Plan National STAMEENA PID2022-139099NB-I00 project funded by MCIN/AEI/10.13039/501100011033 and by the “European Union NextGenerationEU/PRTR" (PRTR-C17.I1), FPI); QUANTERA MAQS PCI2019-111828-2);  QUANTERA DYNAMITE PCI2022-132919 (QuantERA II Programme co-funded by European Union’s Horizon 2020 program under Grant Agreement No 101017733), Ministry of Economic Affairs and Digital Transformation of the Spanish Government through the QUANTUM ENIA project call – Quantum Spain project, and by the European Union through the Recovery, Transformation, and Resilience Plan – NextGenerationEU within the framework of the Digital Spain 2026 Agenda; Fundació Cellex; Fundació Mir-Puig; Generalitat de Catalunya (European Social Fund FEDER and CERCA program, AGAUR Grant No. 2021 SGR 01452, QuantumCAT \ U16-011424, co-funded by ERDF Operational Program of Catalonia 2014-2020); Barcelona Supercomputing Center MareNostrum (FI-2023-1-0013); EU Quantum Flagship (PASQuanS2.1, 101113690); EU Horizon 2020 FET-OPEN OPTOlogic (Grant No 899794); EU Horizon Europe Program (Grant Agreement 101080086 — NeQST), ICFO Internal “QuantumGaudi” project; European Union’s Horizon 2020 program under the Marie Sklodowska-Curie grant agreement No 847648;  “La Caixa” Junior Leaders fellowships, La Caixa” Foundation (ID 100010434): CF/BQ/PR23/11980043. Views and opinions expressed are, however, those of the author(s) only and do not necessarily reflect those of the European Union, European Commission, European Climate, Infrastructure and Environment Executive Agency (CINEA), or any other granting authority.  Neither the European Union nor any granting authority can be held responsible for them. 

\end{acknowledgments}

\bibliography{refs}
\end{document}